\documentclass[pre,twocolumn]{revtex4-2}
\usepackage{calrsfs}
\usepackage{float}
\usepackage{mathtools}
\usepackage{ulem}
\usepackage{amssymb,amsmath}
\usepackage{graphics,epsfig,subfigure}
\usepackage{color,bm,hyperref}
\usepackage{color}
\usepackage{tikz}
\usepackage{yfonts}
\usetikzlibrary{shapes.geometric,arrows}
\usetikzlibrary{decorations.markings}
\DeclareMathAlphabet{\pazocal}{OMS}{zplm}{m}{n}
\begin{document}
\title{Tight-binding model subject to conditional resets at random times}

\author{Anish Acharya}\email{anish.acharya@tifr.res.in}
\affiliation{Department of Theoretical Physics, Tata Institute of Fundamental Research, Homi Bhabha Road,
Mumbai, 400005, India}

\author{Shamik Gupta}\email{shamik.gupta@theory.tifr.res.in}
\affiliation{Department of Theoretical Physics, Tata Institute of Fundamental Research, Homi Bhabha Road,
Mumbai, 400005, India}
\pacs{}

\begin{abstract}
We investigate the dynamics of a quantum system subjected to a time-dependent and conditional resetting protocol. Namely, we ask: what happens when the unitary evolution of the system is repeatedly interrupted at random time instants with an instantaneous reset to a specified set of reset configurations taking place with a probability that depends on the current configuration of the system at the instant of reset? Analyzing the protocol in the framework of the so-called tight-binding model describing the hopping of a quantum particle to nearest-neighbour sites in a one-dimensional open lattice, we obtain analytical results for the probability of finding the particle on the different sites of the lattice. We explore a variety of dynamical scenarios, including the one in which the resetting time intervals are sampled from an exponential as well as from a power-law distribution, and a set-up that includes a Floquet-type Hamiltonian involving an external periodic forcing. Under exponential resetting, and in both presence and absence of the external forcing, the system relaxes to a stationary state characterized by localization of the particle around the reset sites. The choice of the reset sites plays a defining role in dictating the relative probability of finding the particle at the reset sites as well as in determining the overall spatial profile of the site-occupation probability. Indeed, a simple choice can be engineered that makes the spatial profile highly asymmetric even when the bare dynamics does not involve the effect of any bias. Furthermore, analyzing the case of power-law resetting serves to demonstrate that the attainment of the stationary state in this quantum problem is not always evident and depends crucially on whether the distribution of reset time intervals has a finite or an infinite mean.  
\end{abstract}
\maketitle

\section{Introduction}
Stochastic resetting in classical and quantum systems has attracted a lot of attention in recent years owing to the wide range of physical processes that it models, but also owing to the variety of static and dynamic consequences that it may generate~\cite{PhysRevLett.106.160601}. The basic framework involves a system that is very general in every aspect, namely, it could be either classical or quantum, could be either single or many-particle, and could be evolving in time either deterministically or stochastically. Resetting refers to repeated and sudden interruptions of the inherent dynamics of the system with a dynamics that in essence competes with it, and it is this interplay that results in rich and intriguing long-time observable properties of the system. Examples are myriad, and a rather comprehensive discussion of the various examples may be found in the recent reviews on the subject, Refs.~\cite{evans2020stochastic,gupta2022stochastic,nagar2023stochastic}.    

In this work, our focus is on quantum systems. Before delving into our work, it is pertinent to refer to some relevant studies addressing resetting in quantum systems. Earlier works have studied differences in resetting-induced effects in quantum integrable and non-integrable systems~\cite{PhysRevB.98.104309}, quantum coherence and dynamical aspects such as
purity and fidelity in closed quantum systems~\cite{Sevilla_2023}, dynamics of a qubit in presence of detectors, revealing a correspondence between stochastic wave-function dynamics and the underlying resetting process~\cite{Dubey_2023}, and how the reset algorithm along with ballistic propagation facilitates quantum-search processes~\cite{Yin_2023}. In a context similar to the last work, Ref.~\cite{yin2023instability} has reported on the optimal restart times for a quantum first-hitting-time problem. In a different context, it has been shown how the eigenvalue spectrum of a Markovian generator gets shifted in presence of stochastic resetting, with applications to resetting in quantum processes that exhibit metastability~\cite{Rose_2018}. There has been an upsurge of interest in understanding the behaviour of entanglement entropy and correlation functions in resetting processes. For example, Ref.~\cite{PhysRevB.105.L241114} has examined entanglement dynamics in quantum many-body systems allowing for a quasiparticle description and in the presence of stochastic resetting, unveiling interesting entanglement scaling and phase transitions. A recent study~\cite{kulkarni2023generating} has implemented stochastic resetting in closed quantum systems, quantifying the effects in terms of the von Neumann entropy, fidelity, and concurrence. Reference~\cite{Magoni_2022} has unveiled long-range quantum and classical correlations in a non-interacting quantum spin system. In another context, Ref.~\cite{Perfetto_2022} has explored how stochastic resetting may be exploited as a tool to study the thermodynamics of
quantum-jump trajectories in quantum-optical systems, while Ref.~\cite{rieracampeny2021measurementinduced} discusses resetting processes in quantum systems with quantum-collapse considered akin to a measurement. There has been a recent focus on ground state preparation from frustration-free Hamiltonians using resetting techniques; for example, in this regard, Ref.~\cite{puente2023quantum} addresses the ample scope of periodic resetting by invoking ancillary degrees of freedom in quantum systems.

In the above backdrop, our motivation in this work is to study the effects of a time-dependent conditional resetting protocol in quantum systems. To this end, we focus on the paradigmatic tight-binding model (TBM), which models the unitary quantum dynamics of a single particle hopping between the nearest-neighbour sites of a one-dimensional open lattice. In this framework, we implement a protocol of resetting, whereby the particle resets at random times to multiple reset sites with a probability that depends on the current location of the particle. This latter feature makes the reset probability explicitly time-dependent and the resulting resetting protocol a time-dependent one conditioned on the location of the particle at the instant of reset. Specifically, our protocol involves repeated occurrences of a pair of key events: unitary evolution for random times, and instantaneous conditional resetting of the TBM particle to the specified reset sites with a probability that depends on the location of the particle at the instant of reset. Our implementation of the time-dependent reset protocol is inspired by the framework proposed in Ref.~\cite{PhysrevB.104.L180302}, wherein the system, which is a paradigmatic quantum model of interacting spins, undergoes a reset based on the measured magnetization of the system at the instant of reset. In the context of the TBM, a recent work~\cite{lami2023continuously} has addressed the issue of the dynamics undergoing unitary evolution and continuous monitoring, studying in particular the probability distribution of the expectation value of a given observable over the possible quantum trajectories.

Two remarks are in order. Firstly, it is important to recognize that the aforementioned resetting protocol deviates from the conventional approach of time-independent single-shot reset studied in the literature, which involves resetting the system at random times to a predetermined configuration with a probability that is constant and independent of the underlying dynamics (unconditional resetting). By contrast, in our studied protocol, the reset probability being a function of the current location of the particle is explicitly dependent on the underlying reset-free dynamics of the system and is thus explicitly time dependent. Secondly, time-dependent resetting, albeit of a very different sort, has been explored in the context of classical systems in Ref.~\cite{pal2016diffusion}, in which the authors have derived a sufficient condition on the rate of the resetting process for the existence of a stationary state. This reference however serves to show that existence of a stationary state under time-dependent resetting is not always guaranteed, and one has to stipulate conditions on the resetting protocol so that a stationary state exists.

Let us note down what new aspects are to be expected in implementing conditional resetting in quantum systems as opposed to classical systems. The classical equivalent of the tight-binding dynamics will be a classical random walk, albeit in continuous time. In the classical setting, the particle or the random walker in one realization of the dynamics may be found at a given time instant on one and only one site with probability one. Of course, the site label changes from one realization to another of the dynamics. In contrast, in the quantum setting, the particle at a given instant and in one realization of the dynamics may be found simultaneously on many different sites with probabilities that are not equal in general. Therefore, in the latter setting, we may implement at any time instant $t$ a reset to different sites conditioned on the probability at time $t$ for the particle to be found in different subsets of the lattice sites. A protocol constructed in the same vein and when implemented in the classical setting cannot render the resetting move a conditional one of the sort we are considering. This is because the probability for the particle to be found in different subsets of the lattice sites is zero for all but one site, with the probability being unity for this one site.

In the setting proposed within the TBM for studying conditional resetting, we address the following question in a variety of dynamical scenarios: what is the probability of finding the particle at site $m$ at time $t>0$, given that the particle was initially at site $n_0$? As regards the dynamical scenarios, we will first consider the time intervals between successive reset events to be random variables chosen independently from an exponential distribution. In the course of this paper, we will show how the choice of the initial position of the particle and reset site locations will play important roles in illustrating the nature of the site-occupation probability. The main results that emerge are (i) that the dynamics at long times relaxes to a stationary state with the particle localized around the reset sites, which is characterized by time-independent values of the site-occupation probability, (ii) that the site-occupation probability around the two reset sites may or may not be equal to one another, depending on how the reset sites are distributed in space; in the case when the probabilities are unequal, the particle is more likely to be found around one set of reset sites than the rest, implying thereby a reset-induced effective drift in the particle motion despite the fact that the underlying TBM dynamics takes place in the absence of any bias on the particle motion. Our aforementioned analysis of exponential resetting extends even to the case in which the TBM system is subject to a periodic forcing, which renders the underlying Hamiltonian time dependent. Our analytical approach is versatile enough in tacking efficiently and allowing to derive exact results for the case of both the time-independent and the time-dependent TBM system. To provide a counter to the naive expectation that our protocol of conditional resetting always leads to a stationary state, we next study the case of the time intervals between successive resets chosen from a power-law distribution. In this case, for low values of the exponent characterizing the power law, we show that resetting is unable to arrest the ever-spreading site-occupation probability of the TBM particle in time, thereby preventing the system to relax to a stationary state with the particle localized around the reset sites. The mentioned example thus provides an antithesis to the expectation that resetting inherently leads to stationary states. 

We now come to a discussion of why a conditional resetting of the sort we are exploring here may prove useful in developing a better understanding of open quantum systems, and what new physics we may hope to learn from the study. To this end, we first recognize that the TBM  undergoing unitary dynamics that is repeatedly interrupted at random times with stochastic resetting is an example of a more general set-up of a quantum system whose unitary evolution is interspersed with non-unitary interactions (modelled in our case by stochastic resetting) at random times. From the protocol of stochastic resetting, it is evident that the information content of the system, which would otherwise have been preserved had the dynamics been purely unitary, gets lost each time a reset happens: unitary dynamics leads to coherence, while reset events result in decoherence. Indeed, the information on the site-occupation probability  of the TBM particle gets lost every time a reset takes place, such that at just the following time instant, the particle is to be found only at the reset locations and not at any other locations. In this sense, the dynamics models an open quantum system, namely, a system that is not isolated and thus evolving according to a unitary dynamics, but rather one which is coupled to an external environment (namely, the apparatus that implements the stochastic resets). From the foregoing, it is evident that our dynamical set-up of the TBM subject to stochastic resetting involves an interplay of coherence and decoherence, and hence, we may already expect non-trivial consequences arising from this interplay.

The time evolution of the density operator of an isolated quantum system follows the unitary dynamics dictated by the well-known quantum Liouville equation or the von Neumann equation (in this work, we will set the Planck's constant to unity),
\begin{align}
\frac{\mathrm{d}\rho(t)}{\mathrm{d}t}=-\mathrm{i}\mathcal{L}(t)\rho(t),
\end{align}
with $\mathcal{L}(t)$ being the Liouville operator defined as $\mathcal{L}(t)\rho(t)=[H(t),\rho(t)]$, and $H(t)$ being the Hamiltonian (in general, time dependent) of the system. By contrast, the density operator of an open quantum system, under the assumption of being weakly coupled to a Markovian environment, evolves according to a non-unitary dynamics given by the so-called Lindblad equation~\cite{breuer2002theory},
\begin{align} 
\frac{\mathrm{d}\rho(t)}{\mathrm{d}t}=-\mathrm{i}\mathcal{L}(t)\rho(t)+\gamma\left(O\rho(t) O^\dagger - \frac{1}{2}\{O^\dagger O,\rho(t)\}\right),
\label{eq:Lindblad}
\end{align}
where for two operators $A,B$, we have $\{A,B\}\equiv AB+BA$, and where the operator $O$, called the jump operator, describes the dynamics of interaction of the system with the environment, and the constant $\gamma\ge 0$ describes the rate at which the system interacts with the environment. For the case of stochastic resetting of the TBM that is unconditional and happens at a constant rate $\lambda$, it has been shown in Ref.~\cite{Debraj:2022-2} (and which will be elaborated further in Section~\ref{sec:general-set-up}) that resetting at a given instant of time is implemented by the action of a superoperator $T$ on the instantaneous density operator of the system, with $T$ being time independent. In this case, the density operator of the system at any time $t$, when averaged over different realizations of the unitary dynamics of the TBM interspersed with stochastic resets at random times, evolves as~\cite{Debraj:2022-2}
\begin{align}
ˇ\frac{\mathrm{d}\overline{\rho}(t)}{\mathrm{d}t}=-\mathrm{i}\mathcal{L}(t)\overline{\rho}(t)+\lambda T\overline{\rho}(t)-\lambda \overline{\rho}(t),
\label{eq:Lindblad-1}
\end{align}
so that on comparing with Eq.~\eqref{eq:Lindblad}, one obtains $\gamma=\lambda$ and $O\rho O^\dagger=T\rho$. As is evident from the dynamics~\eqref{eq:Lindblad-1}, knowing the average density operator at time $t$ suffices to know its form at a later time $t+\mathrm{d}t$, thus implying Markovian evolution of the average density operator. As shown in Refs.~\cite{Debraj:2022,Debraj:2022-2} and will be further demonstrated in this paper, it is this average density operator which contains all the relevant information on the site-occupation probability of the TBM particle. The consequences of the dynamics in dictating the form of this probability of the TBM particle have been explored in Refs.~\cite{Debraj:2022,Debraj:2022-2}. In implementing the conditional reset explored in this work, we have a time-dependent $T$, that is, $T=T(t)$. We will show in Section~\ref{sec:general-set-up} and Appendix~\ref{sec:app0} that in contrast to the Markovian evolution~\eqref{eq:Lindblad-1}, the average density operator has an evolution that is explicitly non-Markovian and hence more general: we have 
\begin{align}
&\frac{\mathrm{d}}{\mathrm{d} t}\overline{\rho}(t)=-\mathrm{i} \mathcal{L}(t)\overline{\rho}(t)+ \lambda \sum_{\alpha=1}^{\infty} H_{\alpha}(t)-\lambda \overline{\rho}(t),
\label{eq:Lindblad-ours}
\end{align}
with
\begin{align}
&H_1(t) \equiv T(t)~ \mathrm{e}^{-\lambda t}~ \mathrm{e}_+^{-\mathrm{i}\int_0^{t} \mathrm{d}t'~\mathcal{L}(t')} \rho(0), \\
&H_\alpha(t) \equiv \lambda\int_{0}^{t} \mathrm{d}{t_{\alpha-1}}  T(t-t_{\alpha-1}) \mathrm{e}^{-\lambda(t-t_{\alpha-1})}\nonumber \\
&~~~~~~~~\times \mathrm{e}_+^{-\mathrm{i}\int_{t_{\alpha-1}}^{t} \mathrm{d}t'~\mathcal{L}(t')} H_{\alpha-1}(t_{\alpha-1});~\alpha\ge 2.
\end{align}
Indeed, the presence of the term $\lambda \sum_{\alpha=1}^{\infty} H_{\alpha}(t)$ in Eq.~\eqref{eq:Lindblad-ours} implies that knowing $\overline{\rho}(t)$ does not suffice to know the average density operator at a later instant, and in fact, one has to know the full evolution of the system up to time $t$ (history-dependent evolution) to obtain the behaviour at a later time. Thus, our protocol of conditional resetting arms us with a framework in the domain of open quantum systems to explore the consequences of a non-Markovian evolution of the density operator, beyond the more usual Markovian evolution studied in the literature. A non-Markovian evolution, being dependent on the history of evolution, is evidently more non-trivial than Markovian evolution. Yet, it is remarkable that we are able to derive in this paper exact and explicit analytical results for quantities of direct physical relevance such as the site-occupation probability of the TBM particle. The new physics that we may hope to learn from our study is explicit demonstration of the effects of non-Markovian evolution in the context of open quantum systems.

The paper is organized as follows. In Section~\ref{sec:general-set-up}, we provide a general framework to study quantum evolution interspersed with instantaneous yet  time-dependent non-unitary interactions at random time intervals. The non-unitary interactions are modelled in terms of a time-dependent interaction superoperator acting on the density operator of the system. The developed formalism is next applied to the tight-binding model, a brief summary of whose salient dynamical features is presented in Section~\ref{sec: Application of the protocol to Tight Binding Model}. Section~\ref{sec: exponential resetting} forms the core of the paper, wherein we present analytical results, suitably validated by numerical results, on how exponential resetting alters the behavior of the TBM. We discuss in turn our computation of the site-occupation probability of the TBM particle for various choices of reset sites and initial location of the TBM particle, highlighting in particular how and why such choices manifest in the stationary behavior of the site-occupation probability. We also discuss the effects of turning on an external periodic forcing that acts on the TBM particle. Section~\ref{sec:power-law} discusses the drastic change in the behavior of the site-occupation probability when the resetting time intervals are sampled from a power-law distribution, emphasizing, in particular, the conditions required for the emergence of a stationary state under our protocol of conditional resetting. The paper ends with conclusions in Section~\ref{sec:conclusion}. Some of the technical details of our analytic computation are collected in the three appendices.

\section{The general framework}
\label{sec:general-set-up}

Consider a generic quantum system described by a time-dependent Hamiltonian $H(t)$. The dynamics of the system involves unitary evolution dictated by $H(t)$, which is interspersed at random times with instantaneous interactions of the system with the external environment or a measuring device. These instantaneous interactions induce non-unitarity in the evolution of the system. We take these interactions to be time-dependent. Thus, our present set-up is a non-trivial generalization of the one introduced and studied in Refs.~\cite{Debraj:2022} and \cite{Debraj:2022-2}. 

\begin{figure}[H]
\includegraphics[width=1\linewidth]{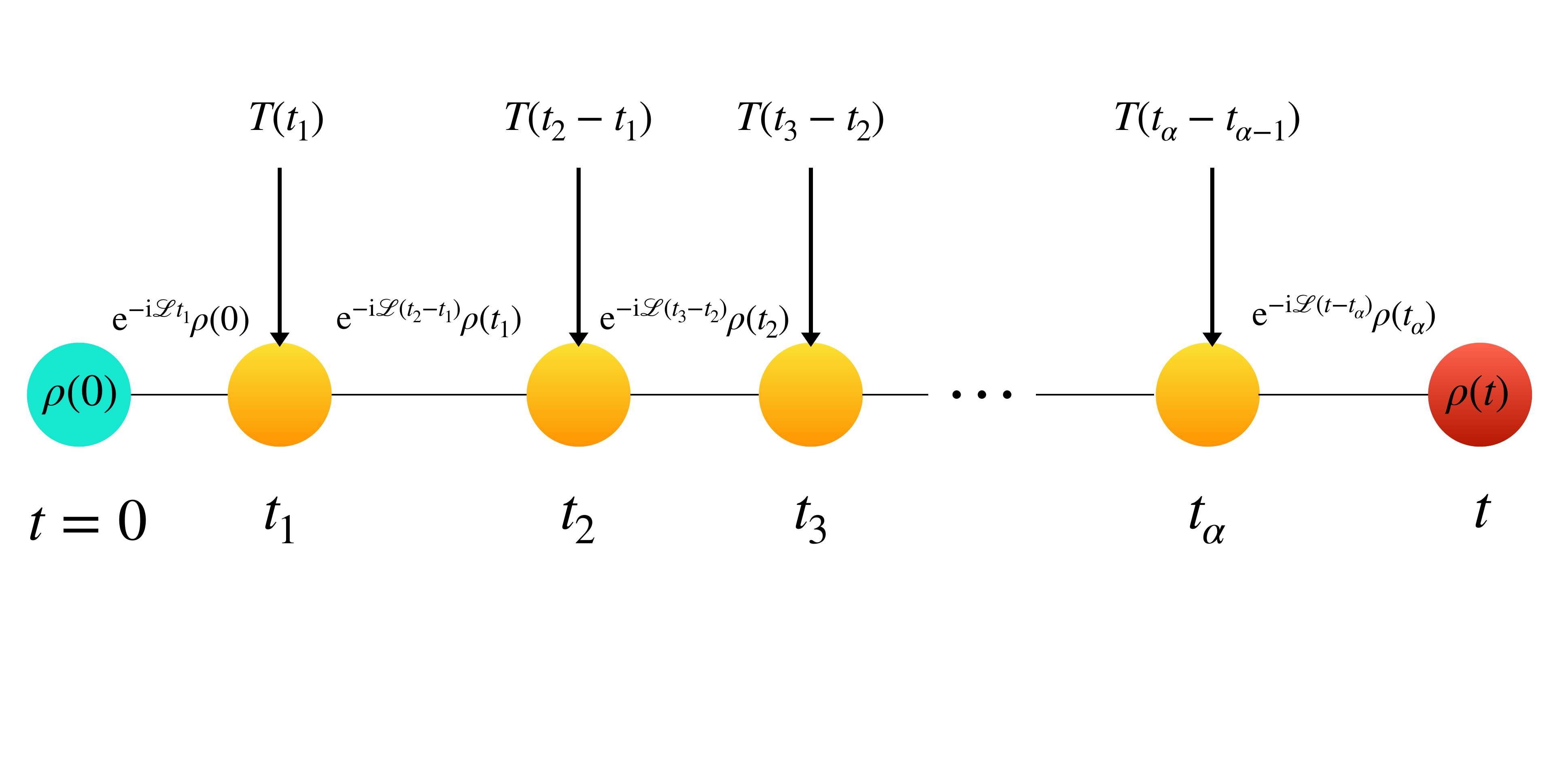}
\caption{For a generic quantum system undergoing unitary evolution interspersed at random times with instantaneous and non-unitary time-dependent interactions, the figure shows a schematic representation of the dynamical evolution of the density operator for the case of a time-independent Hamiltonian. Here, $\rho(0)$ is the initial density operator, $\rho(t)$ is the density operator at the time instant $t$ of interest, the $t_\alpha$'s are the instances of non-unitary interactions modelled by the interaction operator $T(t_\alpha - t_{\alpha-1})$, while $\mathcal{L}$ is the Liouville operator describing the unitary evolution.}
        \label{fig:density-operator}
\end{figure}

The evolution of the density operator of the system for a fixed time duration $[0,t]$ involves the following: A unitary evolution for a random time interval $[0,t_1]$ is followed by an instantaneous non-unitary interaction at the time instant $t_1$ modelled by a time-dependent interaction superoperator $T(t_1)$.  As compared to an ordinary operator that acts on a state vector to yield a state vector, a superoperator acts on an ordinary operator to give another ordinary operator~\cite{Debraj:2022}. The instantaneous interaction at time instant $t_1$ is followed by a unitary evolution for a random time interval $[t_1,t_2]$, which is followed by a second instantaneous non-unitary interaction at $t_2$ described by the superoperator $T(t_2-t_1)$, and so on. The interaction superoperator at any instant of time is thus a function of the time elapsed since the previous interaction. Over the time interval $[0, t]$, a realization of the evolution of the system involves a certain $\alpha \ge 0$ number of non-unitary interactions at random time instants $t_1,t_2,\ldots,t_\alpha$, with unitary evolution for the intermediate time intervals $t_1-0, t_2-t_1, t_3-t_2, t_4-t_3, \ldots,t_\alpha-t_{\alpha-1},t-t_\alpha$. We take the time intervals $\tau_{\alpha'+1} \equiv t_{\alpha'+1} - t_{\alpha'};
~\alpha'=0,1,2,\ldots,\alpha-1;~t_0 =0$ between successive interactions to be independent and identically-distributed (i.i.d.) random variables sampled from a common distribution $p(\tau)$. Averaging over different realizations of the dynamics, the average density operator at time $t$ is evidently given by~\cite{Debraj:2022-2} 
\begin{widetext}
   \begin{eqnarray}
    &&\overline{\rho}(t)=U(t)\rho(0);\nonumber \\
    &&U(t) \equiv \sum_{\alpha=0}^\infty \int_0^t \mathrm{d}t_\alpha~ \int_0^{t_\alpha} \mathrm{d}t_{\alpha-1}\ldots\int_0^{t_3} \mathrm{d}t_2 \int_0^{t_2} \mathrm{d}t_1 F(t-t_\alpha)\;\mathrm{e}_+^{-\mathrm{i}\int_{t_{\alpha}}^{t} \mathrm{d} t' \mathcal{L}(t')}\;T(t_\alpha-t_{\alpha-1})\;p(t_\alpha-t_{\alpha-1})\;\nonumber \\
    &&~~~~~~~\times\; \mathrm{e}_+^{-\mathrm{i}\int_{t_{\alpha-1}}^{t_\alpha}\mathrm{d} t'\mathcal{L}(t')}\ldots T(t_2-t_1)\;p(t_2-t_1)\;\mathrm{e}_+^{-\mathrm{i}\int_{t_1}^{t_2}\mathrm{d} t'\mathcal{L}(t')}\;T(t_1)\;p(t_1)\;\mathrm{e}_+^{-\mathrm{i}\int_0^{t_1} \mathrm{d} t'\mathcal{L}(t')},
    \label{eq:average-rhot_tbh}
    \end{eqnarray} 
\end{widetext}
where $\rho(0)$ is the initial density operator and $U(t)$ denotes the non-unitary time-evolution superoperator. Moreover, because $H(t)$ at two different times may not commute with each other, we have invoked time ordering in writing down the exponential factors in the above equation; the minus and plus subscripts on the exponential indicate negative (i.e., the latest time to the right) and positive (i.e., the latest time to the left) time ordering, respectively~\cite{Debraj:2022-2}. The operator $\mathrm{e}_+^{-\mathrm{i} \int_{t'}^t\mathrm{d} \tau \mathcal{L}(\tau)}$, with $\mathcal{L}$ being the Liouville operator, describes unitary evolution and is defined by its operation on a density operator as~\cite{Debraj:2022-2}
\begin{align}
    \rho(t)&=\mathrm{e}_+^{-\mathrm{i} \int_{t'}^t\mathrm{d} \tau~ \mathcal{L}(\tau)}\rho(t'<t)\nonumber\\
    &=\mathrm{e}_+^{-\mathrm{i} \int_{t'}^t\mathrm{d}\tau~ H(\tau)}\rho(t'<t)~\mathrm{e}_-^{\mathrm{i} \int_{t'}^t\mathrm{d}\tau~ H(\tau)}.
\end{align}
Note that for the case of a time-independent $H$, no time ordering is required; consequently, the above equation takes a simplified form: $\rho(t)=\mathrm{e}^{-\mathrm{i}\mathcal{L}(t-t')}\rho(t'<t)\nonumber=\mathrm{e}^{-\mathrm{i} H(t-t')}\rho(t'<t)~\mathrm{e}^{\mathrm{i}H(t-t')}$. 
The quantity $F(t)$ in Eq.~\eqref{eq:average-rhot_tbh} denotes the probability for no interaction during time duration $t$; given the distribution $p(\tau)$, one obtains $F(t)$ as  $F(t)=\int_t^\infty \mathrm{d}\tau~p(\tau)$. Figure~\ref{fig:density-operator} shows a schematic depiction of the time evolution of the density operator for one realization of the dynamics and for the case of a time-independent Hamiltonian.

Although at first glance, the expression for $\overline{\rho}(t)$ in Eq.~\eqref{eq:average-rhot_tbh} appears formidable for analysis, it does take a simpler form in the Laplace-transformed space. Given the Laplace convolution theorem that the Laplace transform (denoted by the operator $\pazocal{L}$) of a convolution $f(t)*g(t) \equiv \int_0^t \mathrm{d}t'~ f(t') g(t-t')$ of two functions $f(t)$ and $g(t)$ equals  $\pazocal{L}[f(t)*g(t)]=\pazocal{L}[f(t)]\pazocal{L}[g(t)]$, Eq.~\eqref{eq:average-rhot_tbh} in the Laplace domain $s$ writes as 
\begin{align}
\widetilde{\overline{\rho}}(s)=\widetilde{U}(s)\rho(0),
\label{eq:rhotildes}
\end{align}
where we have 
\begin{align}
    &\widetilde{U}(s)\nonumber\\
    &= \pazocal{L}[F(t)\;\mathrm{e}_+^{-\mathrm{i}\int_0^t\mathrm{d} t' \mathcal{L}(t')}]\sum_{\alpha=0}^\infty\Bigl(\pazocal{L}[T(t)p(t)\; \mathrm{e}_+^{-\mathrm{i} \int_0^t\mathrm{d} t' \mathcal{L}(t')}]\Bigr)^\alpha \nonumber\\
    &=  \dfrac{\pazocal{L}[F(t) \mathrm{e}_+^{-\mathrm{i}\int_0^t \mathrm{d} t' \mathcal{L}(t')}]}{I-\pazocal{L}\left[T(t)p(t)\;\mathrm{e}_+^{-\mathrm{i}\int_0^t \mathrm{d} t' \mathcal{L}(t')}\right]},
    \label{eq:U_tilde}
\end{align}
with $\widetilde{U}(s)$ being the Laplace transform of the superoperator $U(t)$ and $I$ denoting the identity operator. Before proceeding further, let us remark that Eq.~\eqref{eq:U_tilde} is very general in that it applies to any Hamiltonian (time-dependent or time-independent), to any form of the interaction operator $T(t)$ (time-dependent or time-independent), and to any distribution $p(\tau)$ of the time interval between successive interactions. 

Let us now specialize to exponential $p(\tau)$, namely, \begin{align}
p(\tau)=\lambda \mathrm{e}^{-\lambda \tau},
\label{eq: exp_dist}
\end{align}
yielding 
\begin{align}
F(t)= \mathrm{e}^{-\lambda t}.
\label{eq:Ft}
\end{align}
Here, the parameter $\lambda>0$ denotes the interaction rate or the probability of interaction per unit time. The quantity $1/\lambda$ yields the average time between two successive interactions. In this case, we show in Appendix~\ref{sec:app0} how differentiating Eq.~\eqref{eq:average-rhot_tbh} with respect to time yields the Lindblad-like equation for the time evolution of the average density operator as given by Eq.~\eqref{eq:Lindblad-ours}.

Now, for the case of exponential $p(\tau)$, using Eqs.~\eqref{eq: exp_dist} and~\eqref{eq:Ft}, the numerator and the denominator of Eq.~\eqref{eq:U_tilde} are respectively given by
\begin{align}
    \pazocal{L}[F(t)\;\mathrm{e}_+^{-\mathrm{i}\int_0^t \mathrm{d} t' \mathcal{L}(t')}]
    &=\int_0^{\infty} \mathrm{d} t\;\mathrm{e}^{-(s+\lambda)t}\mathrm{e}_+^{-\mathrm{i}\int_0^t \mathrm{d} t' \mathcal{L}(t')}\nonumber\\
    &\equiv\widetilde{U}_0(s)\label{eq: U_0(s)}
\end{align}
and
\begin{align}
      &\pazocal{L}[T(t)p(t)\; \mathrm{e}_+^{-\mathrm{i}\int_0^t\mathrm{d} t'\mathcal{L}(t')}]\nonumber\\
      &=\lambda\int_0^\infty \mathrm{d}t\;\mathrm{e}^{-(s+\lambda) t}\; T(t)\; \mathrm{e}_+^{- \mathrm{i} \int_0^t \mathrm{d} t'\mathcal{L}(t')} \nonumber\\
      &=\lambda\int_0^\infty \mathrm{d}t\;\mathrm{e}^{-s't}\; T'(t) \nonumber\\
      &=\lambda\; \widetilde{T}'(s'),
\label{eq: T_tilde_s}
\end{align}
where we have defined $s'\equiv(s+\lambda)$ and $ T'(t)\equiv T(t)\;\mathrm{e}_+^{-\mathrm{i}\int_0^t \mathrm{d} t' \mathcal{L}(t')}$. Equation~\eqref{eq:U_tilde} may therefore be written as
\begin{align}
\widetilde{U}(s)= \dfrac{\widetilde{U}_0(s)}{I-\lambda \widetilde{T}^{'}(s') };
{\label{eq: U_tilde_unexpanded}}\end{align}
expanding the right hand side as a series in $\lambda$ by using the  operator identity $(A-B)^{-1}= A^{-1}+A^{-1}B(A-B)^{-1}$ leads to the result
\begin{align}
    &\widetilde{U}(s)=\widetilde{U}_0(s)\sum_{\alpha=0}^{\infty}\lambda^\alpha  
    (\widetilde{T}'(s'))^{\alpha}.
{\label{eq: U_tilde_S}}\end{align}

Note that the operator $\widetilde{U}_0(s)$ is the Laplace transform of the product of the interaction-free unitary-evolution operator $U_0(t) \equiv \mathrm{e}_+^{-\mathrm{i}\int_0^t \mathrm{d} t' \mathcal{L}(t')}$ and the factor $F(t)=\mathrm{e}^{-\lambda t}$. In contrast to the series expansion outlined in Ref.~\cite{Debraj:2022}, the present expansion~\eqref{eq: U_tilde_S} offers a greater degree of generality by factoring in the time dependence of the interaction operator. In the following, we will apply the developed formalism, first to the case of a time-independent Hamiltonian and then to the time-dependent case. As mentioned previously, the model we will consider is the so-called tight-binding model, which plays a paradigmatic role in studies in solid-state physics, e.g., in modeling nano wires~\cite{PhysrevB.34.3625,dattagupta2022driven}.
 
\section{The Tight-Binding Model (TBM)}
\label{sec: Application of the protocol to Tight Binding Model}
The tight-binding model (TBM) in one dimension is a simple and representative model for studying quantum dynamics. The model involves a single quantum particle hopping between the nearest-neighbour sites of a one-dimensional lattice with open boundaries. We take the lattice to be extending from $-\infty$ to $\infty$, with sites labeled by the index $n$ and the site $n=0$ being the origin. The dynamics of the system is described by the time-independent Hamiltonian
\begin{align}
H=-\dfrac{\gamma}{2}\sum_{n=-\infty}^{\infty}(|n\rangle \langle n+1 | + |n+1\rangle \langle n | ),
\label{eq:H}
\end{align}
where $\gamma > 0$ is the strength of nearest-neighbour hopping, and the Wannier states $|n\rangle$ stand for the state of the particle located on site $n$. In terms of the operators 
\begin{align}
K\equiv \sum_{n=-\infty}^{\infty} | n \rangle \langle n+1|,~~K^{\dagger}\equiv \sum_{n=-\infty}^{\infty} | n+1 \rangle \langle n |,
\label{eq:KKdagger}
\end{align}
the Hamiltonian may be expressed as
\begin{align}
H=-\frac{\gamma}{2}(K+ K^{\dagger} ),
\label{eq:H-KKdagger}
\end{align}
the advantage being that $[K, K^{\dagger}]=0$ allows diagonalization of the Hamiltonian in terms of the basis formed by the simultaneous eigenstates of $K$ and $K^{\dagger}$ called the Bloch states. The latter states are defined in the momentum space reciprocal to the space spanned by the Wannier states and obey 
\begin{align}
K| k \rangle= \mathrm{e}^{-\mathrm{i} k} | k \rangle,
\label{eq:K-eigenvalue}
\end{align}
where we have
\begin{align}
| k \rangle \equiv \frac{1}{\sqrt {2\pi}}\sum_{n=-\infty} ^\infty \mathrm{e}^{-\mathrm{i} n k} | n \rangle,~~ |n\rangle \equiv \frac{1}{\sqrt{2\pi}}\int_{-\pi}^\pi \mathrm{d}k~\mathrm{e}^{\mathrm{i}nk}|k\rangle.
\label{eq:Bloch-states}
\end{align}

Let us now ask the question: Given that the particle at the initial instant $t=0$ was on site $n_0$, what is the probability $P_m(t)$ of finding it on an arbitrary site $m$ at a later time $t>0$? In terms of the Wannier states, we have $P_m(t)=|\langle m | \mathrm{e}^{-\mathrm{i} Ht} |n_0 \rangle|^2$. A straightforward calculation of this probability involving the Bloch states can be carried out in terms of the density operator $\rho(t)$ of the system, which evolves in times as $\rho(t)=\mathrm{e}^{-\mathrm{i}Ht}\rho(0)\mathrm{e}^{\mathrm{i}Ht}$\label{eq: time_evolution for time independent H}. Here, the initial density operator is given by $\rho(0)=|n_0\rangle \langle n_0|$, while in terms of $\rho(t)$, the probability $P_m(t)$ is evidently given by $P_m(t)=\langle m|\rho(t)|m\rangle= \langle m|\mathrm{e}^{-\mathrm{i} H t } | n_0 \rangle \langle n_0|\mathrm{e}^{\mathrm{i} H t }| m\rangle$. Using Eq.~\eqref{eq:Bloch-states}, one obtains 
\begin{equation} 
 \mathrm{e}^{\mathrm{i}H t} |n\rangle = \frac{1}{\sqrt{2\pi}}\int_{-\pi}^{\pi} \mathrm{d}k~\mathrm{e}^{\mathrm{i} nk -\mathrm{i} \gamma t\cos{k}} |k\rangle,
 \label{eq:bare-TBM-result}
 \end{equation}
  which finally yields~\cite{Debraj:2022}
\begin{align}
      P_m(t)&= \frac{1}{(2 \pi)^2} \int_{-\pi}^{\pi}\int_{-\pi}^{\pi} \mathrm{d}k\;\mathrm{d}k'~\mathrm{e}^{\mathrm{i} (m-n_0)(k-k') - \mathrm{i}\Gamma_{k k'}t}\nonumber\\
      &=J^2_{|m-n_0|}(\gamma t)\label{free_prop},
\end{align}
where we have
\begin{align}
    \Gamma_{\scriptscriptstyle k k'}\equiv\gamma(\cos k'-\cos k),
\end{align}
and $J_m(x)$ is the Bessel function of the first kind of order $m$ with the property $J_{-m}(x)=(-1)^mJ_m(x)$~\cite{NIST:DLMF}. Using the identity $\sum_{m=-\infty}^\infty J_m^2(x)=1~~\forall~x$~\cite{NIST:DLMF}, it is checked that as desired, the probability $P_m(t)$ is properly normalized to unity: $\sum_{m=-\infty}^{\infty} P_m(t)=1$. From its very definition, it follows that the quantity $P_m(t)$ may be interpreted as the propagator of the TBM, with the quantity $1/\gamma$ evidently being the inherent dynamical time scale of the system.

Having obtained the site-occupation probability in Eq.~\eqref{free_prop}, one may obtain the mean displacement from the initial location as $\langle m -n_0\rangle=\sum_{m=-\infty}^\infty (m-n_0)P_m(t)=0$, where we have used the fact that $P_{m}(t)=P_{-m}(t)$. The next quantity of interest is the mean-squared displacement (MSD) from the initial location, obtained as
\begin{align}
S(t)&\equiv \sum_{m=-\infty}^{\infty} (m-n_0)^2~P_m(t)=\frac{\gamma^2 t^2}{2},\label{eq: MSD_TBM_noreset} 
\end{align}
where we have used the result $\sum_{m=-\infty}^\infty m^2 J^2_m(x)=x^2/2$ \cite{Mathematica}. Some features are evident from the results~\eqref{free_prop} and~\eqref{eq: MSD_TBM_noreset}. The bare TBM dynamics does not have a stationary state, which is owing to the fact that the TBM particle moves on an open lattice, and so spreads out to a larger number of lattice sites with respect to its initial location $n_0$ as time progresses. Concomitantly, the MSD grows forever as a function of time. The quadratic growth of the MSD with time may be contrasted with the corresponding result in classical diffusion in which the MSD grows linearly with time. In the following section, we will unveil how these features get drastically modified when the TBM is subject to time-dependent interactions of a particular sort, namely, that modeling conditional stochastic resetting. 

\begin{figure}
\includegraphics[width=0.8\linewidth]{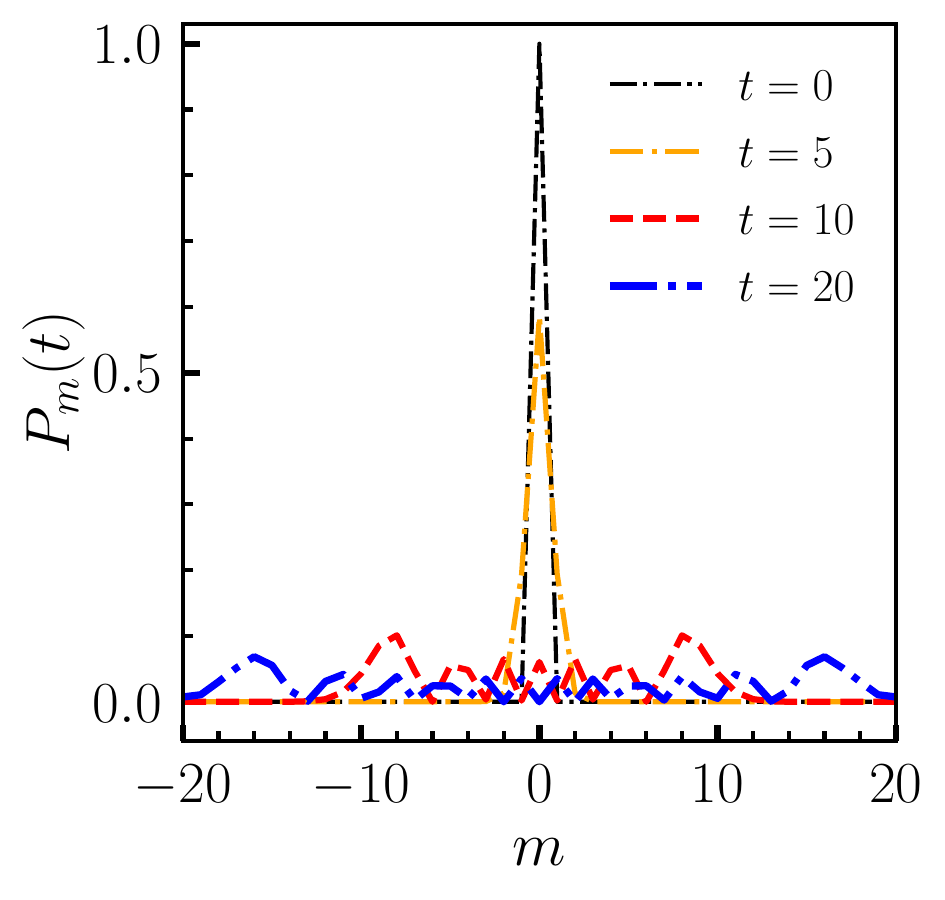}
\caption{For the bare TBM undergoing unitary evolution according to the Hamiltonian~\eqref{eq:H}, the figure shows the site-occupation probability $P_m(t)$, given by Eq.~\eqref{free_prop}, at four time instances. The particle is taken to be initially located on site $n_0=0$, while the parameter $\gamma$ has the value $\gamma=1$. One may observe the spreading of the probability with the passage of time, implying the absence of any stationary state with the particle localized in space.}
        \label{fig: 2}
\end{figure}
\section{TBM subject to conditional resetting at exponentially-distributed times}
\label{sec: exponential resetting}
In this work, we will choose a particular form of interaction that implements an instantaneous reset of the TBM particle to a specific set of sites. Specifically, the system evolves via the repetitive occurrence of the following pair of events: unitary evolution for a random time governed by the Hamiltonian~\eqref{eq:H}, followed by an instantaneous relocation of the particle to two possible reset locations with a probability that depends on the current location of the particle at the time instant of reset. Thus, our set-up deviates significantly from earlier work on stochastic resets in quantum systems, specifically, in the context of the TBM, in which relocation to a fixed site with a rate that is constant and independent of the current location of the particle has been pursued~\cite{Debraj:2022, Debraj:2022-2}. The TBM (and related systems) subjected to projective measurements has been extensively pursued while addressing the issue of detection problems corresponding to a quantum particle arriving at a chosen set of sites \cite{PhysRevA.91.062115, dhar2015quantum, Friedman_2016, Friedman_2017, Liu_2022, PhysRevA.103.022222, PhysRevA.103.032221, PhysRevResearch.2.043107, PhysRevLett.120.040502, thiel2019quantum}. Our work is distinct in that it deals with a theme that has not been considered in the above references, namely, the protocol of conditional stochastic resetting that we now define.
\begin{figure}
\includegraphics[width=1\linewidth]{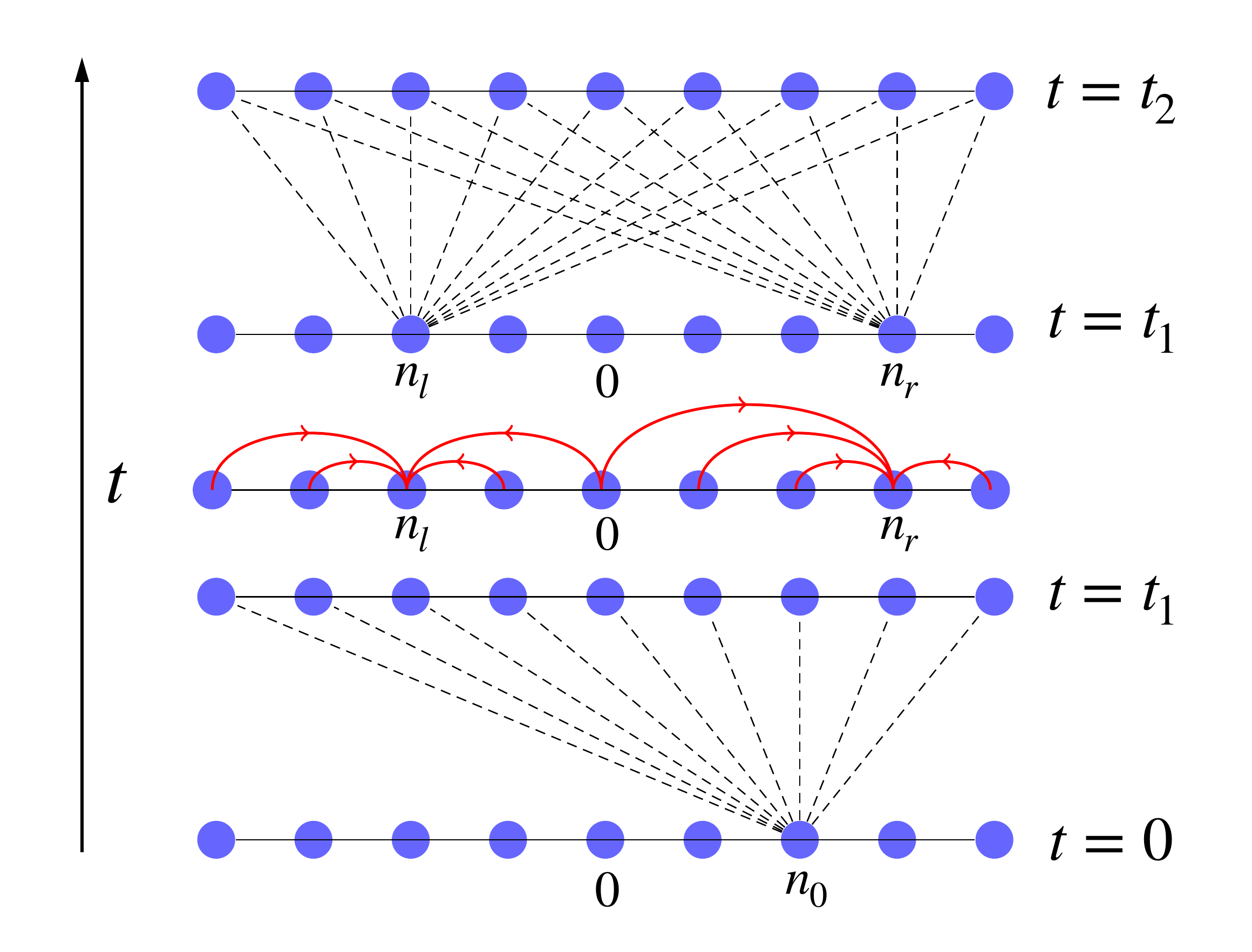}
\caption{Illustration of the conditional resetting of the TBM particle. The figure shows the evolution of the system as time progresses while starting with the TBM particle on site $n_0$ on the lattice at time $t=0$. The system evolves unitarily in time for a random time duration $t_1$, which is indicated in the figure by the dashed lines; at time instant $t_1$, the particle has different probabilities to be found on different sites on the lattice; subsequently, (i.e., after a random time interval $\tau_1$ with respect to the initial time $t=0$), the system undergoes an instantaneous conditional reset to sites $n_r$ and $n_l$ with probabilities given by Eq.~\eqref{eq:first-reset}. Namely, the probability to reset to site $n_r$ (respectively, to site $n_l$) is given by the probability at time $t_1$ for the particle to be found on any site to the right (respectively, to the left) of the origin ($n=0$) and half the probability for the particle to be at the origin at time $t_1$. These reset events are shown by the red curves in the figure. Subsequent to the first reset at time instant $t_1$, the system undergoes unitary evolution for the random time interval $\tau_2=t_2-t_1$, following which there is the second reset to sites $n_r$ and $n_l$ (not shown in the figure) with probabilities given by Eq.~\eqref{eq:second-reset}. The evolution proceeds through such alternating events of unitary evolution and instantaneous conditional resetting.}.
\label{fig: 3}
\end{figure}

While starting with the particle on a given site $n_0$ at time $t=0$, our protocol of conditional reset conditioned on the current location of the particle is as follows. The particle evolves unitarily for a random time $\tau_1$ distributed according to the exponential~\eqref{eq: exp_dist} and then undergoes an instantaneous reset. The location the particle resets to depends on its location just prior to the reset, namely, the location at time instant $t_1=\tau_1$. Specifically, on being located to the right (respectively, to the left) of the origin $n=0$, the particle resets to a given site $n_r$ to the right (respectively, to a given site $n_l$ to the left) of the origin. On the other hand, if the particle at time $t_1=\tau_1$ is located at the origin, it resets with an equal probability of $1/2$ to the sites $n_r$ and $n_l$, respectively. Let us denote the probability to reset to the sites $n_r$ and $n_l$ by $\Omega_{n_r}^{n_0}(\gamma\tau_1)$ and $\Omega_{n_l}^{n_0}(\gamma\tau_1)$, respectively. Noting that at time $t_1=\tau_1$, the probability for the particle to be located to the right and to the left of the origin are respectively given by  $\sum_{j=1}^{\infty}|\langle j | \mathrm{e}^{-\mathrm{i} H \tau_1} | n_0 \rangle |^2$ and $\sum_{j=-1}^{-\infty}|\langle j | \mathrm{e}^{-\mathrm{i} H \tau_1} | n_0 \rangle |^2$, while that for the particle to be located at the origin is $|\langle 0 | \mathrm{e}^{-\mathrm{i} H \tau_1} | n_0 \rangle |^2$, we obtain 
\begin{align}  
&\Omega_{n_r}^{n_0}(\gamma\tau_1)=\sum_{j=1}^{\infty}|\langle j | \mathrm{e}^{-\mathrm{i} H \tau_1} |n_0 \rangle |^2 +\frac{1}{2}|\langle 0 | \mathrm{e}^{-\mathrm{i} H \tau_1} | n_0\rangle |^2, \nonumber\\
   \label{eq:Qn0}\\
   &\Omega_{n_l}^{n_0}(\gamma\tau_1)=\sum_{j=-1}^{-\infty}|\langle j | \mathrm{e}^{-\mathrm{i} H \tau_1} | n_0 \rangle |^2 +\frac{1}{2}\big|\langle 0 | \mathrm{e}^{-\mathrm{i} H \tau_1} | n_0 \rangle |^2. \nonumber 
   \end{align}
Employing Eq.~\eqref{free_prop}, the above expressions of probabilities rewrite as
 \begin{align}
 &\Omega_{n_r}^{n_0}(\gamma \tau_1)=\sum_{j=1}^{\infty} J_{|j-n_0|}^2(\gamma \tau_1) +\frac{1}{2}J_{|n_0|}^2(\gamma \tau_1), \nonumber \\ \label{eq:first-reset}\\
 &\Omega_{n_l}^{n_0}(\gamma \tau_1)=\sum_{j=-1}^{-\infty} J_{|j-n_0|}^2(\gamma \tau_1) +\frac{1}{2}J_{|n_0|}^2(\gamma \tau_1). \nonumber 
   \end{align}
Note that using $\sum_{m=-\infty}^\infty J_m^2(x)=1~\forall~x$, we obtain for arbitrary $t$ the normalization 
\begin{align}
\Omega_{n_r}^{n_0}(\gamma t)+\Omega_{n_l}^{n_0}(\gamma t) = 1~\forall~n_0.
\label{eq:norm-Q}
\end{align}
 
The first reset at time $t_1$ is followed by unitary evolution for a random time interval $\tau_2$ distributed according to the distribution~\eqref{eq: exp_dist}, which is followed by an instantaneous reset at time $t_2=t_1+\tau_2$. In accordance with our protocol, the reset may happen to either the site $n_r$ or the site $n_l$, depending on whether the particle at time $t_2$ is located, respectively, to the right and to the left of the origin. In other words, the reset at time $t_2$ to sites $n_r$ and $n_l$ takes place with respective probabilities $\Theta_{n_r}(\gamma\tau_1,\gamma\tau_2)$ and $\Theta_{n_l}(\gamma\tau_1,\gamma\tau_2)$, with 
 \begin{align}
 &\Theta_{n_r}(\gamma\tau_1,\gamma\tau_2)=\Omega_{n_r}^{n_l}(\gamma \tau_2) \Omega_{n_l}^{n_0}( \gamma \tau_1)+\Omega_{n_r}^{n_r}(\gamma \tau_2)\Omega_{n_r}^{n_0}(\gamma \tau_1),\nonumber \\
 \label{eq:second-reset} \\
 &\Theta_{n_l}(\gamma\tau_1,\gamma\tau_2)= \Omega_{n_l}^{n_l}(\gamma \tau_2) \Omega_{n_l}^{n_0}(\gamma\tau_1) + \Omega_{n_l}^{n_r}(\gamma \tau_2)\Omega_{n_r}^{n_0}(\gamma\tau_1). \nonumber \end{align}
 Continuing it this way, the probabilities for the next reset after a random interval $\tau_3$ (i.e., at time instant $t_3=t_2+\tau_3$) are given by
 \begin{align}
 &\Theta_{n_r}(\gamma\tau_1,\gamma\tau_2,\gamma\tau_3)=\Omega_{n_r}^{n_l}(\gamma\tau_3)\Theta_{n_l}(\gamma\tau_1,\gamma\tau_2)\nonumber\\
 &+\Omega_{n_r}^{n_r}(\gamma \tau_3)\Theta_{n_r}(\gamma\tau_1,\gamma\tau_2),\nonumber \\
\label{eq:third-reset}\\
 &\Theta_{n_l}(\gamma\tau_1,\gamma\tau_2,\gamma\tau_3)
 = \Omega_{n_l}^{n_l}(\gamma\tau_3)\Theta_{n_l}(\gamma\tau_1,\gamma\tau_2)\nonumber\\
 &+ \Omega_{n_l}^{n_r}(\gamma\tau_3)\Theta_{n_r}(\gamma\tau_1,\gamma\tau_2).\nonumber 
\end{align}
Proceeding in this manner, we obtain the iterative and conditional structure of the reset probabilities. Note that we have the conservation of probability:
\begin{align}
&\Theta_{n_r}(\{\gamma\tau_\alpha\})+ \Theta_{n_l}(\{\gamma\tau_\alpha\})=1;\nonumber \\
&\Theta_{n_r}(\gamma\tau_1)=\Omega_{n_r}^{n_0}(\gamma\tau_1),~\Theta_{n_l}(\gamma\tau_1)=\Omega_{n_l}^{n_0}(\gamma\tau_1).
\label{eq:Q-norm}
\end{align}
A realization of the dynamics for a fixed time $t$ comprises the aforementioned alternating sequence of unitary evolution for a random time, which is followed by an instantaneous reset to sites $n_l$ and $n_r$ with probabilities that depend on the full history of time evolution up to the instant of reset, see Eqs.~\eqref{eq:second-reset} and~\eqref{eq:third-reset}. 

We would like to now study our model by employing the general framework developed in Section~\ref{sec:general-set-up}. Our object of study is the probability 
\begin{align}
\overline{P}_m(t)=\langle m|\overline{\rho}(t)|m\rangle
\end{align}
for the TBM particle to be found on site $m$ at time $t$, given that it was on site $n_0$ at time $t=0$; the overbar denotes averaging over different realizations of the dynamics of unitary evolution interspersed with instantaneous conditional resets at random times. Working in the Laplace space $s$, the mentioned probability reads as \begin{align}
\widetilde{\overline{P}}_m(s)=\langle m|\widetilde{\overline{\rho}}(s)|m\rangle,   
\end{align} whose evaluation requires that of the quantity $\widetilde{\overline{\rho}}(s)$ given by Eqs.~\eqref{eq:rhotildes} and~\eqref{eq: U_tilde_S}. 

In order to compute $\widetilde{\overline{\rho}}(s)$, we need to first write down an explicit expression of the time-dependent interaction superoperator $T(t)$ that implements our protocol of conditional resetting. In this respect, we note that the superoperator $T(t)$ projects the density operator prior to a reset into a density operator characterizing a mixed state. Let $\rho_{\mp}(\tau_1)$ denote the density operator just before and just after the first reset taking place at time instant $t_1=\tau_1$. We evidently have $\rho_+(\tau_1)=\Omega_{n_r}^{n_0}(\gamma \tau_1)|n_r\rangle \langle n_r|+\Omega_{n_l}^{n_0}(\gamma\tau_1)|n_l\rangle \langle n_l|$, while by definition, we have $\rho_+(\tau_1)
    =T(\tau_1)\rho_{-}(\tau_1)$, with $\mathrm{Tr}(\rho_\pm(\tau_1))=1$. That the density operator just before a reset has a trace equal to unity follows from the fact that the trace of the density operator just after the previous reset has trace equal to unity and the fact that the time evolution between two resets is unitary and thus trace-preserving. Consequently, we must have $T(\tau_1)$ of the form 
\begin{align}
     &(n_1 n_2| T(\tau_1)| n_3 n_4)\nonumber\\
     &=\delta_{n_3 n_4}(\delta_{n_1 n_r}\delta_{n_2 n_r}  \Omega_{n_r}^{n_0}(\gamma \tau_1)+\delta_{n_1 n_l}\delta_{n_2 n_l}  \Omega_{n_l}^{n_0}(\gamma \tau_1));\label{eq: T1(t)}
 \end{align}
 indeed, we have 
\begin{align}
     &\langle n_1|T(\tau_1)\rho_{-}(\tau_1)|n_2\rangle\nonumber\\
     &=\sum_{n_3 n_4} ( n_1 n_2 |T(\tau_1)| n_3 n_4 ) \langle n_3|\rho_{-}(\tau_1)|n_4\rangle\nonumber\\
     &=\sum_{n_3 n_4}  \delta_{n_3 n_4}(\delta_{n_1 n_r}\delta_{n_2 n_r} \Omega_{n_r}^{n_0}(\gamma \tau_1)+\delta_{n_1 n_l}\delta_{n_2 n_l} \Omega_{n_l}^{n_0}(\gamma \tau_1))\nonumber \\
     &\times \langle n_3|\rho_-(\tau_1)|n_4\rangle\nonumber\\
     &=\delta_{n_1 n_r}\delta_{n_2 n_r} \Omega_{n_r}^{n_0}(\gamma \tau_1)+\delta_{n_1 n_l}\delta_{n_2 n_l} \Omega_{n_l}^{n_0}(\gamma \tau_1)\nonumber \\
     &=\langle n_1|\rho_+(\tau_1)|n_2\rangle,
     \label{eq:checking-Tt1}
 \end{align}
where in arriving at the first step, we have used Eq.~\eqref{eq: super operation matrix element}, while to obtain the second last step, we have used $\sum_{n_1n_2}  \delta_{n_1 n_2}\langle n_1|\rho_-(\tau_1)|n_2\rangle=\mathrm{Tr}(\rho_-(\tau_1))=1$. Proceeding as above, one may straightforwardly check that the superoperator $T(\tau_2)$ for the second reset at time $t_2=t_1+\tau_2$ must satisfy 
\begin{align}
     (n_1 n_2| T(\tau_2)|n_3 n_4)
     &=\delta_{n_3 n_4}(\delta_{n_1 n_r}\delta_{n_2 n_r} \Theta_{n_r}(\gamma\tau_1,\gamma\tau_2))\nonumber\\
     &+\delta_{n_1 n_l}\delta_{n_2 n_l} \Theta_{n_l}(\gamma\tau_1,\gamma\tau_2)),\label{eq: T2(t)}
 \end{align}
 ensuring that 
 \begin{align}
    &\rho_+(\tau_2)
    =T(\tau_2)\rho_-(\tau_2)\nonumber\\
    &= \Theta_{n_r}(\gamma\tau_1,\gamma\tau_2)|n_r\rangle \langle n_r|+\Theta_{n_l}(\gamma\tau_1,\gamma\tau_2)|n_l\rangle \langle n_l|,
\end{align}
and that 
\begin{align}
      &(n_1 n_2| T(\tau_3)|n_3 n_4)=\delta_{n_3 n_4}(\delta_{n_1 n_r}\delta_{n_2 n_r} \Theta_{n_r}(\gamma\tau_1,\gamma\tau_2,\gamma\tau_3)\nonumber\\
     &+\delta_{n_1 n_l}\delta_{n_2 n_l} \Theta_{n_l}(\gamma\tau_1,\gamma\tau_2,\gamma\tau_3)),\label{eq: T3(t)}
 \end{align}
 so that 
 \begin{align}
    &\rho_+(\tau_3)
    =T(\tau_3)\rho_-(\tau_3)\nonumber\\
    &= \Theta_{n_r}(\gamma\tau_1,\gamma\tau_2,\gamma\tau_3)|n_r\rangle \langle n_r| +\Theta_{n_l}(\gamma\tau_1,\gamma\tau_2,\gamma\tau_3)|n_l\rangle \langle n_l|.
\end{align}

Armed with the knowledge that we now have of the form of the interaction superoperator $T(\tau)$ for any $\tau$, we now proceed to obtain explicit results for the probability $\overline{P}_m(t)$ for the TBM particle to be on site $m$ at time $t$, given its initial location $n_0$. 

\subsection{Computation of site-occupation probability} \label{sec: analysis for time-independent TBC}
In order to compute the site-occupation probability $\overline{P}_m(t)$, we start with the corresponding expression in the Laplace domain $s$, obtained combining Eqs.~\eqref{eq:rhotildes} and~\eqref{eq: U_tilde_S} as
\begin{align}        \widetilde{\overline{P}}_m(s)
        &=\langle m |\widetilde{U}_0(s)\sum_{\alpha=0}^{\infty}\lambda^\alpha  (\widetilde{T}'(s'))^{\alpha} \rho(0)| m\rangle \nonumber\\
&=\sum_{\alpha=0}^{\infty}\widetilde{\overline{P}}_m^{(\alpha)}(s),
\label{eq: probability series terms}
\end{align}
 where $\widetilde{\overline{P}}_m^{(\alpha)}(s)\equiv \langle m|\widetilde{U}_0(s)\lambda^\alpha  (\widetilde{T}'(s'))^{\alpha}\rho(0)|m\rangle $ is the contribution involving $\alpha$ number of resets to the quantity $\widetilde{\overline{P}}_m(s)$. The corresponding equation in the time domain has the obvious form
\begin{align}        \overline{P}_m(t)
&=\sum_{\alpha=0}^{\infty}\overline{P}_m^{(\alpha)}(t).
\label{eq: probability series terms-time}
\end{align}

The TBM Hamiltonian~\eqref{eq:H} being time-independent guarantees the commutativity of the Hamiltonian at distinct times and so no time ordering is required to be invoked in our analysis. Consequently, the quantities $\widetilde{U}_0(s)$ and  $\widetilde{T}'(s')$ in Eqs.~\eqref{eq: U_0(s)} and~\eqref{eq: T_tilde_s}, respectively, are modified to
\begin{align}
&\widetilde{U}_0(s)\equiv\int_0^{\infty} \mathrm{d} t\;\mathrm{e}^{-(s+\lambda)t}\;\mathrm{e}^{-\mathrm{i}\mathcal{L}t}, \nonumber \\ \\
    &\widetilde{T}'(s')=
    \int_0^\infty \mathrm{d}t\;\mathrm{e}^{-(s+\lambda) t}\; T(t)\; \mathrm{e}^{-\mathrm{i} \mathcal{L}t}.\nonumber
\end{align} 

 To proceed requires calculation of the matrix elements of $\widetilde{U}_0(s)$ and $\widetilde{T}'(s')$; the former computation has been previously presented in Ref.~\cite{Debraj:2022} and is recalled here for use in subsequent analysis. We have
 \begin{align}
   &(n_1' n_2'| \widetilde{U}_0(s)|n_1 n_2)\nonumber\\
   &=\int_0^{\infty} \mathrm{d}\tau\;\mathrm{e}^{-(s+\lambda) \tau} \langle n_1'| \mathrm{e}^{-\mathrm{i} H \tau}|n_1 \rangle \langle n_2| \mathrm{e}^{\mathrm{i} H \tau}|n_2'\rangle\nonumber\\
   &=\frac{1}{(2 \pi)^2}\int_{-\pi}^{\pi}\int_{-\pi}^{\pi} \mathrm{d}k\;\mathrm{d}k' \dfrac{ \: \mathrm{e}^{\mathrm{i}k(n_1'-n_1)-\mathrm{i}k'(n_2'-n_2)}}{(s+\lambda)I+ \mathrm{i} \Gamma_{k k'}}\label{eq:U_0}.
 \end{align}
 Here, in the first step, we have used Eq.~\eqref{eq:iLt-elements}, while the second step is obtained on use of Eqs.~\eqref{eq:bare-TBM-result} and~\eqref{eq:Bloch-states}. Moreover, we have used the result that one has $\pazocal{L}(\mathrm{e}^{-\alpha t})=1/(s+\alpha)$.
 Let us then evaluate the first term of the series given by Eq.~{\eqref{eq: probability series terms}}; we have
\begin{align}
\widetilde{\overline{P}}_m^{(0)}(s)
      &= \langle m |\widetilde{U}_0(s) \rho(0)|m \rangle \nonumber\\
      &=\sum_{n_1,n_2}(m m| \widetilde{U}_0(s)| n_1 n_2) \langle n_1 | \rho(0)| n_2 \rangle\nonumber \\
      &=(m m| \widetilde{U}_0(s)| n_0 n_0),
\end{align}
where we have used $\rho(0)=|n_0\rangle\langle n_0|$ and Eq.~\eqref{eq: super operation matrix element}. Using Eq.~\eqref{eq:U_0}, we get
\begin{align}
&\widetilde{\overline{P}}_m^{(0)}(s)=\frac{1}{(2 \pi)^2}\int_{-\pi}^{\pi}\int_{-\pi}^{\pi} \mathrm{d}k\;\mathrm{d}k' \dfrac{ \: \mathrm{e}^{\mathrm{i}(k-k')(m-n_0)}}{(s+\lambda)I+ \mathrm{i} \Gamma_{k k'}},
\end{align}
which using the Laplace convolution theorem leads to the corresponding result in the time domain, as
\begin{align}
    \overline{P}_m^{(0)}(t)
&=\dfrac{\mathrm{e}^{-\lambda t}}{( 2 \pi)^2}\int_{-\pi}^\pi  \int_{-\pi}^\pi \mathrm{d}k \;  \mathrm{d}k' \mathrm{e}^{\mathrm{i}(m-n_0)(k-k')} \mathrm{e}^{- \mathrm{i}\Gamma_{k k'} t}\nonumber\\
&=\mathrm{e}^{-\lambda t}J^2_{|m-n_0|}(\gamma t),
\label{eq:Pm0t}
\end{align}
 where in the second line, we have used Eq.~\eqref{free_prop}. 

 Next, we may calculate the second term of the series in Eq.~{\eqref{eq: probability series terms}} by using the matrix element of the superoperator $T$ given by Eq.~\eqref{eq: T1(t)}. Relegating the details of the derivation to Appendix~\ref{sec:app2}, we mention here the final result that reads 
\begin{align}
\overline{P}_m^{(1)}(t)
   &= \lambda \mathrm{e}^{-\lambda t}\int_0^t \mathrm{d}t_1  ( J^2_{|m-n_r|}(\gamma (t-t_1) ) \Omega_{n_r}^{n_0}(\gamma t_1)\nonumber\\
   &+  J^2_{|m-n_l|}(\gamma (t-t_1))
   \Omega_{n_l}^{n_0}(\gamma t_1)).\label{eq: Pm1(t)}
\end{align}

In a similar manner, the third term of the series in Eq.~\eqref{eq: probability series terms} may be evaluated
to obtain in the time domain that (see Appendix~\ref{sec:app3}): 
\begin{widetext} 
\begin{align}
\overline{P}_m^{(2)}(t)
    &=\lambda^2 \mathrm{e}^{-\lambda t } \int_0^t \mathrm{d} t_2 \int_0^{t_2} \mathrm{d} t_1 \Big[J_{|m-n_r|}^2(\gamma (t-t_2))\left(\Omega_{n_r}^{n_r}(\gamma(t_2-t_1)) 
     \Omega_{n_r}^{n_0}(\gamma t_1)+\Omega_{n_r}^{n_l}(\gamma(t_2-t_1))\Omega_{n_l}^{n_0}(\gamma t_1)\right)\nonumber\\
    &+J_{|m-n_l|}^2(\gamma (t-t_2) ) \left(\Omega_{n_l}^{n_r}(\gamma(t_2-t_1))\Omega_{n_r}^{n_0}(\gamma t_1)+\Omega_{n_l}^{n_l}(\gamma(t_2-t_1))\Omega_{n_l}^{n_0}(\gamma t_1)\right)\Big],\label{eq: Pm2(t)}
   \end{align}
   \end{widetext}
and the $\alpha$-th ($\alpha \ge 1$) term of the series in Eq.~\eqref{eq: probability series terms} turns out to be
\begin{widetext}
\begin{align}
    \overline{P}_m^{(\alpha)}(t)
    &=\lambda^{\alpha} \mathrm{e}^{-\lambda t } \Big[\smallint_0^t   \mathrm{d} t_{\alpha}\smallint_0^{t_\alpha}\mathrm{d} t_{\alpha-1}\ldots \smallint_0^{t_3}\mathrm{d} t_2 \smallint_0^{t_2}\mathrm{d} t_1  \Big[ J_{|m-n_r|}^2(\gamma (t-t_{\alpha}))\hspace{-0.5cm}\sum_{\mu_1,\mu_2,\ldots,\mu_\alpha}\hspace{-0.4cm}\Omega_{n_r}^{\mu_\alpha}(\gamma( t_\alpha-t_{\alpha-1}))\ldots \Omega_{\mu_2}^{\mu_1}(\gamma(t_2-t_1))\Omega_{\mu_1}^{n_0}(\gamma t_1)\nonumber\\
    &+J_{|m-n_l|}^2(\gamma (t-t_\alpha))\hspace{-0.5cm} \sum_{\mu_1,\mu_2,\ldots,\mu_\alpha}\hspace{-0.4cm}\Omega_{n_l}^{\mu_\alpha}(\gamma( t_\alpha-t_{\alpha-1}))\ldots \Omega_{\mu_2}^{\mu_1}(\gamma(t_2-t_1))\Omega_{\mu_1}^{n_0}(\gamma t_1) \Big]  \label{eq: general Pm alpha},
   \end{align}
\end{widetext}
where the $\mu_\alpha$'s can be either $n_r$ or $n_l$. We see from the above expression that $\overline{P}_m^{(\alpha)}(t)$ is obtained from contributions of all possible trajectories connecting the initial site $n_0$ and the site $m$ of interest, via the reset sites $n_r$ and $n_l$, with resetting taking place at times $t_1,t_2,\ldots,t_\alpha$. 

Thus far we have been very general as regards the choice of the initial location $n_0$. From Eq.~\eqref{eq: general Pm alpha}, it is evident that for the number of resets $\alpha \geq 2$, the complexity of the nested integral in the expression of $\overline{P}_m^{(\alpha)}(t)$ continues to grow with $\alpha$ owing to the iterative structure of the interaction superoperator $T$. This feature renders the resulting analytical expression more challenging to handle in the time domain. Nevertheless, we will particularly be interested in the stationary-state properties of the dynamics, and, in particular, in the behavior of $\overline{P}_m(t)$ as $t \to \infty$. We show below that despite the mentioned analytical complexity, explicit results may be derived for the stationary state for arbitrary choice of $n_0$. We treat first the simpler case of reset sites equidistant with respect to the initial location at origin, and then move on to discuss the more general case.

\subsubsection{Reset sites equidistant from the initial location taken to be at the origin} \label{subsec: initial location at origin with equidistant reset locations}

Let us consider a special case: the reset sites are equidistant from the initial location of the TBM particle that we take to be at the origin $n_0=0$; thus, we have $n_l=-n_r$. From Eq.~\eqref{eq:first-reset}, we note on setting $n_0=0$ that $\Omega_{n_r}^{n_0}(\gamma\tau_1)=\Omega_{n_l}^{n_0}(\gamma\tau_1)$, so that Eq.~\eqref{eq:norm-Q} implies then that $\Omega_{n_r}^{n_0}(\gamma\tau_1)=\Omega_{n_l}^{n_0}(\gamma\tau_1)=1/2$. Moreover, we have $\Omega_{n_r}^{n_r}(\gamma\tau_2)=\Omega_{n_l}^{n_l}(\gamma\tau_2)$ and $\Omega_{n_l}^{n_r}(\gamma\tau_2)=\Omega_{n_r}^{n_l}(\gamma\tau_2)$, so that Eq.~\eqref{eq:second-reset}, together with Eq.~\eqref{eq:Q-norm}, implies that $\Theta_{n_l}(\gamma\tau_1,\gamma\tau_2)= \Theta_{n_r}(\gamma\tau_1,\gamma\tau_2)=1/2$. Arguing in a similar manner leads to the result  $\Theta_{n_l}(\{\gamma\tau_\alpha\})= \Theta_{n_r}(\{\gamma\tau_\alpha\}) = 1/2$. 
From Eq.~\eqref{eq: Pm1(t)}, we have  
\begin{align}
&\overline{P}_m^{(1)}(t)\nonumber\\
   &= \dfrac{\lambda}{2} \;\mathrm{e}^{-\lambda t}\int_0^t \mathrm{d}t_1  ( J^2_{|m-n_r|}(\gamma (t-t_1) )+  J^2_{|m-n_l|}(\gamma (t-t_1))).\label{eq: Pm1(t)_mid}    
\end{align}

On the other hand, the ${\alpha}$-th $(\alpha\geq 2)$ term, Eq.~\eqref{eq: general Pm alpha}, involving $\alpha$ number of resetting events is obtained as 
\begin{align}
\overline{P}_m^{(\alpha)}(t)
    &= \frac{\lambda^{\alpha}}{2} \mathrm{e}^{-\lambda t}\int_{0}^{t} \mathrm{d}t'\frac{(t')^{\alpha-1}}{(\alpha-1)!}\nonumber\\
    &\times(J_{|m-n_r|}^2(\gamma (t-t'))+ J_{|m-n_l|}^2(\gamma (t-t'))).\label{eq: Pm_kt_midpoint}
\end{align}
Consequently, we obtain on using the above results and Eq.~\eqref{eq:Pm0t} with $n_0=0$ in Eq.~\eqref{eq: probability series terms-time} the desired probability of finding the particle at site $m$ at time $t$ as
\begin{align}
     &\overline{P}_m (t)\nonumber\\
     &= \mathrm{e}^{-\lambda t} J_{|m|}^2(\gamma t)\nonumber\\ &+ \dfrac{\lambda}{2}\mathrm{e}^{-\lambda t}\int_0^t \mathrm{d}t_1 ( J^2_{|m-n_r|}(\gamma (t-t_1))+J^2_{|m-n_l|}(\gamma (t-t_1)))\nonumber\\
     &+\dfrac{ \lambda}{2} \mathrm{e}^{-\lambda t}\int_{0}^{t} \mathrm{d}t'  (\mathrm{e}^{-\lambda t'}-1)( J_{|m-n_r|}^2(\gamma (t-t'))\nonumber\\
     &+J_{|m-n_l|}^2 (\gamma (t- t'))),
\end{align}
where the third term is obtained by summing the contribution of $\overline{P}_m^{(\alpha)}(t)$ from $\alpha=2$ to $\alpha=\infty$. The above equation can be further simplified, yielding finally that
\begin{align}
       \overline{P}_m (t)
       &=\mathrm{e}^{-\lambda t} J_{|m|}^2(\gamma t)\nonumber \\ &+ \dfrac{\lambda}{2}\:\int_0^t \mathrm{d}t' \:\mathrm{e}^{-\lambda (t-t')} ( J^2_{|m-n_r|}(\gamma (t-t'))\nonumber\\
       &+ J^2_{|m+n_r|}(\gamma (t-t'))),\label{eq: P_tot(t)_assumption}
\end{align}
where we have used $n_l=-n_r$. Using $\sum_{m=-\infty}^\infty J_m^2(x)=1$, it may be easily checked from the above equation that $\overline{P}_m(t)$ is normalized to unity. The site-occupation probability given by~\eqref{eq: P_tot(t)_assumption} is symmetric under $n_r \to -n_r$, which is a manifestation of the reset sites being equidistant from the origin. An implication is that the TBM particle has equal probability to be found on the two reset sites $n_r$ and $-n_r$.

 Equation~\eqref{eq: P_tot(t)_assumption} may be explained physically as follows: The first term on the right-hand side arises
from those realizations of evolution for time $t$ that did not undergo a single reset since $t=0$ (the
probability for which is $\exp(-\lambda t)$), and consequently, the corresponding contribution to $\overline{P}_m (t)$ is given by the probability of no reset for time $t$ multiplied by the probability $J^2_{|m|}(\gamma t)$ to be on site $m$ at time $t$ in the absence of any reset and while starting from
site $n_0=0$ at time $t = 0$. In order to understand the second term on the right-hand
side, we note that every reset collapses the state of the system to either $|-n_r\rangle$ or $|n_r\rangle$ with an equal probability of $1/2$, and its evolution starts afresh from either the site $-n_r$ or the site $n_r$. Consequently, at time $t$, what
matters is when the last reset took place, and the corresponding contribution to $\overline{P}_m (t)$ is given by the product of the probability $\lambda \mathrm{d}t' \exp(-\lambda (t-t'))$ at time $t$ of the last reset to take place in the
interval $[t',t' -\mathrm{d}t']$, with $t'\in [0, t]$, with the probability in the absence of any reset for
the particle to be on site $m$ resulting from evolution for time duration $(t-t')$ while starting
from site $-n_r$ or site $n_r$. The latter probabilities are given by $J^2_{|m+n_r|}(\gamma (t-t'))$ and $J^2_{|m-n_r|}(\gamma (t-t'))$, respectively. We thus see that although our resetting protocol has an explicit memory of time evolution up to the instant of reset, as implied by Eqs.~\eqref{eq:second-reset} and~\eqref{eq:third-reset}, for the specific case of the initial location of the TBM particle to be at the origin and the two reset locations being equidistant from the origin, what matters in determining the probability $\overline{P}_m (t)$ is the evolution since the last reset and all memories of time evolution previous to the last reset become irrelevant. Equation~\eqref{eq: P_tot(t)_assumption} has an exact correspondence with the last-renewal-equation approach discussed in  Refs.~\cite{PhysRevLett.106.160601}and~\cite{Debraj:2022} to determine reset-induced probability distributions for cases in which the reset protocol is such as not having any memory dependence, that is, the reset at any time instant takes place with a constant probability that is independent of the time evolution up to the instant of reset. 

Next, we obtain the mean displacement from the initial location $n_0=0$, given by $\overline{[m - n_0]}(t)=\sum_{m=-\infty}^\infty m \overline{P}_m(t)$, as
\begin{align}
    \overline{m}(t)=0,
    \label{eq:mean-displacement}
\end{align}
using $\sum_{m=-\infty}^\infty m J^2_m(t)=0$. Let us remark that the mean displacement from the initial location being zero is also observed in the bare TBM dynamics. The reason is the presence of symmetry in the site-occupation probability about the origin in both the cases under consideration. 

On the other hand,  the MSD of the TBM particle from its initial location, defined as
 \begin{align}
     \overline{S}(t) \equiv \sum_{m=-\infty}^{\infty} (m-n_0)^2~\overline{P}_m(t),
     \label{eq:MSD}
 \end{align}
may be evaluated by using Eqs.~\eqref{eq: P_tot(t)_assumption} and~\eqref{eq: MSD_TBM_noreset} as
\begin{align}
    \overline{S}(t)&= \frac{\gamma^2}{\lambda^2}\left(1- \mathrm{e}^{-\lambda t}(1+\lambda t)\right)+(1-\mathrm{e}^{-\lambda  t}) n_r^2.
    \label{eq: MSD_reset_with_assumptions}
\end{align}    

Considering Eq.~\eqref{eq: P_tot(t)_assumption} in the large-$t$ limit yields the stationary-state probability $\overline{P}_m^{\mathrm{st}}$. Clearly, the first term in Eq.~\eqref{eq: P_tot(t)_assumption} decays to zero at large times, and we obtain 
\begin{align}
\overline{P}_m^{\mathrm{st}}= \dfrac{\lambda}{2}\:\int_0^\infty \mathrm{d}t' \:\mathrm{e}^{-\lambda t'} ( J^2_{|m-n_r|}(\gamma t')+ J^2_{|m+n_r|}(\gamma t')).\label{eq:Pmst}
\end{align}
We now use the identity~\cite{gradshteyn2007}
\begin{align}
    \int_0^\infty \hspace{-0.1cm} \mathrm{d}t\;\mathrm{e}
    ^{-a t} J_{\nu}(b t) J_{\nu}(c t) = \frac{1}{\pi (b c)^{\frac{1}{2}}} Q_{\nu-\frac{1}{2} }\left(\frac{a^2+b^2+c^2}{2 b c} \right),\label{eq: Bessel integral}
\end{align}
with the conditions: $\mathrm{Re}[a \pm \mathrm{i}(b+c)]>0$, $c>0$ and $\mathrm{Re}[\nu]>-\frac{1}{2}$. Here,
$Q_{\nu}(\theta)$ is the Legendre function of the second kind, obtained by putting $\mu=0$ in the associated Legendre function of the second kind denoted by $Q^{\mu}_{\nu}(\theta)$, with $\mu$ and $\nu$ being real numbers. The latter function may be related to the Olver's Hypergeometric function 
 $_2 \mathrm{F} _1(a,b;c;z)$ as~\cite{NIST:DLMF}
\begin{align}
    Q^{\mu}_{\nu }(\theta)&=\sqrt{\pi}\; \mathrm{e}^{\mathrm{i}\pi \mu }\frac{\Gamma  (\mu +\nu +1) \left(\theta^2-1\right)^{\mu /2}}{2^{\nu +1}\; \theta^{\mu +\nu +1}}\nonumber\\
    & _2 \mathrm{F} _1\left(\frac{\mu }{2}+\frac{\nu }{2}+1,\frac{\mu }{2}+\frac{\nu }{2}+\frac{1}{2};\nu +\frac{3}{2};\frac{1}{\theta^2}\right),\label{eq: associated legendre}
\end{align}
with $\mu+\nu \ne -1,-2,-3,\ldots$.
Here, $\Gamma(z)$ is the Gamma function. Using Eq.~\eqref{eq: Bessel integral} in Eq.~\eqref{eq:Pmst}, we obtain 
\begin{align}
\overline{P}_m^{\mathrm{st}}= \dfrac{\lambda}{2\pi\gamma} \left( Q_{|m-n_r|-\frac{1}{2} }(\theta)+Q_{|m+n_r|-\frac{1}{2} }(\theta)\right),\label{Eq: Stationary state probability}
\end{align}
with 
\begin{align}
\theta \equiv 1+\frac{\lambda^2}{2 \gamma^2}.\label{eq: def_theta}
\end{align}
 The stationary-state probability~\eqref{Eq: Stationary state probability} offers a stark contrast when viewed against the result obtained in Ref.~\cite{Debraj:2022} due to the presence of two terms accounting for the reset to the sites $n_r$ and $-n_r$, while in the mentioned reference, the TBM particle was considered to be resetting to a single site $\mathcal{N}$. 

As is well known, quantum measurements lead to non-negligible and irreversible interaction between the measuring apparatus and the system of interest, quite unlike classical measurement processes. Indeed, one of the postulates of quantum mechanics is that every measurement involves instantaneous projection of the state of the system onto the quantum state dictated by the measuring apparatus~\cite{cohen1986quantum}. It was shown by Misra and Sudarshan that when such measurements that project the system to the initial state are carried out at regular intervals of time, the system remains frozen in the initial state in the limit in which the measurements are sufficiently frequent. This is called the quantum Zeno effect~\cite{misra1977zeno,georgescu2022quantum}. The Zeno effect has been explored extensively in the literature in a variety of contexts, e.g., in solving optimization problems with multiple arbitrary constraints including inequalities~\cite{herman2023constrained}, in discussing noise correlations in photon polarization~\cite{virzi2022quantum}, in addressing universal control between non-interacting qubits~\cite{blumenthal2022demonstration}, and many others. 
To investigate the Zeno-limit behavior of Eq.~\eqref{Eq: Stationary state probability}, let us consider the limit $\lambda/\gamma \to \infty$. This limit may be achieved by considering at a fixed $\gamma$ the limit $\lambda\to \infty$, which corresponds to making measurements at very frequent time intervals (recall that $1/\lambda$ is the average time interval between two successive resetting), thus conforming to the Zeno set-up. Using the series expression of Olver's hypergeometric function given by~\cite{NIST:DLMF}
\begin{align}
    _2 \mathrm{F}_1\left(a,b;c;z \right)= \frac{1}{\Gamma(a) \Gamma(b)} \sum_{l=0}^{\infty}\frac{\Gamma(a+l) \Gamma(b+l)}{l!~\Gamma (c+l) } z^l ,
\end{align}
 with $|z
 |<1$, one obtains $ \lim_{z\to 0} {}_2 \mathrm{F}_1\left(a,b;c;z \right)=1$. Using this result in Eq.~\eqref{Eq: Stationary state probability} yields 
\begin{align}
    \overline{P}_m^{\mathrm{st}} = \frac{1}{2}( \delta_{m, n_r} + \delta_{m, -n_r}).
    \label{eq:zeno}
\end{align}
In the Zeno limit, the Zeno effect that involves projective measurements to the initial state implies that the system remains frozen in the initial state. By contrast, under conditional resetting of the TBM particle to predefined reset locations that are equidistant from the initial location at the origin, Eq.~\eqref{eq:zeno} implies that the system in the same limit localizes perfectly at these locations, and one has equal probability of finding the particle in the two locations.

In the stationary state, while the mean displacement from the initial location is zero,
\begin{align}
\overline{m}^\mathrm{st}=0, \label{eq:mean-center-stationary}
\end{align}
Eq.~\eqref{eq: MSD_reset_with_assumptions} yields the stationary-state MSD as
\begin{align}
\overline{S}^\mathrm{st}=\frac{\gamma^2}{\lambda^2}+n_r^2.
 \label{eq:MSD-center-stationary}
\end{align}
We now move on to discuss the case in which the reset sites are not equidistant from the initial location, with the latter taken to be arbitrary.

\subsubsection{Arbitrary initial location with reset sites not equidistant from the latter}
We now consider the general case of arbitrary initial location and arbitrary reset sites, namely, ones that are not necessarily equidistant with respect to the initial location. As we will unveil, the consequent memory dependence of the reset probabilities will have a non-trivial effect on the site-occupation probabilities $\overline{P}_m(t)$. To this end, we have on using Eqs.~\eqref{eq: probability series terms-time} and~\eqref{eq:Pm0t} that
\begin{align}
     \overline{P}_m (t)
      &= \mathrm{e}^{-\lambda t} J_{|m-n_0|}^2(\gamma t) + \sum_{\alpha=1}^{\infty} \overline{P}_m^{(\alpha)}(t).\label{alpha series time domain}
 \end{align}
To proceed, we use Eq.~\eqref{eq: general Pm alpha} to write $\overline{P}_m^{(\alpha)}(t)$ in a form similar to Eq.~\eqref{eq: Pm1(t)}, as  
\begin{align}
    \overline{P}_m^{\alpha+1}(t)&=\int_0^t \mathrm{d}t_{\alpha+1} ( J^2_{|m-n_r|}(\gamma(t-t_{\alpha+1})) \mathcal{R}_{n_r}(\gamma t_{\alpha+1})\nonumber\\
    &+J^2_{|m-n_l|}(\gamma(t-t_{\alpha+1})) \mathcal{R}_{n_l}(\gamma t_{\alpha+1}));~~\alpha \ge 1,
    \label{eq:Palpha-recursive}
\end{align}
with 
\begin{align}
\mathcal{R}_{\mu}(\gamma t_{\alpha+1})&\equiv \int_0^{t_{\alpha+1}} \mathrm{d} t_{\alpha} (\Omega_{\mu}^{n_r}(\gamma(t_{\alpha+1}-t_\alpha)) \mathcal{R}_{n_r}(\gamma t_\alpha)\nonumber\\ &+\Omega_{\mu}^{n_l}(\gamma(t_{\alpha+1}-t_\alpha)) \mathcal{R}_{n_l}(\gamma t_\alpha) );~~\alpha \ge 1,
\label{eq:R-recursive}
\end{align}
and $\mathcal{R}_{\mu}(\gamma t_\alpha)$ denoting the effective reset probability for resetting to site $\mu$ (with $\mu$ = $n_l$ or $n_r$) at time instant $t=t_\alpha$.
Equation~\eqref{eq:R-recursive}, with the initial condition
\begin{align}
\mathcal{R}_{\mu}(\gamma t_1)=\Omega_{\mu}^{n_0}(\gamma t_1),
\end{align}
when used in Eq.~\eqref{eq:Palpha-recursive} provides a recursive structure for the quantity $\overline{P}_m^{(\alpha)}(t)$. This structure allows to obtain numerically the quantities $\overline{P}_m^{\alpha+1}(t)$ for $\alpha \ge 1$ and hence the desired probability $\overline{P}_m(t)$. Once the latter is known, the mean-displacement~$\overline{[m-n_0]}(t)$ and the MSD~$\overline{S}(t)$ can also be computed numerically.

It turns out that unlike the case treated in Section~\ref{subsec: initial location at origin with equidistant reset locations}, a closed-form expression for $\overline{P}_m(t)$ cannot be obtained in the present case of general initial location $n_0$ and general reset locations $n_l$ and $n_r$. Yet, for the latter scenario, one obtains quite remarkably exact results for the stationary state $\overline{P}_m^{\mathrm{st}}$, a calculation we turn to in the following. 

Obtaining the stationary state requires evaluating $\overline{P}_m^{(\alpha)}(t)$ and hence $\overline{P}_m(t)$ in the limit $t\to \infty$. Owing to the complicated and nested structure of the integral determining the quantity $\overline{P}_m^{(\alpha)}(t)$, see Eq.~\eqref{eq: general Pm alpha} or Eqs.~\eqref{eq:Palpha-recursive} and~\eqref{eq:R-recursive}, the stationary state may be more easily obtained by considering $\overline{P}_m^{(\alpha)}(t)$ in the Laplace domain $s$. In this case, Eq.~\eqref{alpha series time domain} along with Eq.~\eqref{eq: general Pm alpha} gives
 \begin{align}   
 &\widetilde{\overline{P}}_m(s)=\frac{1}{\pi \gamma}\Bigl(  Q_{|m-n_0|-\frac{1}{2}}\nonumber \\
 &+\lambda~Q_{|m-n_r|-\frac{1}{2}}\bigl((1+\lambda \widetilde{\Omega}_{n_r}^{n_r}+\ldots)\widetilde{\Omega}_{n_r}^ {n_0}+(\lambda\widetilde{ \Omega}_{n_r}^{n_l}+\ldots )\widetilde{\Omega}_{n_l}^{n_0}\bigr)
   \nonumber\\&+\lambda~Q_{|m-n_l|-\frac{1}{2}}\bigl((1+\lambda\widetilde{\Omega}_{n_l}^{n_l}+\ldots)\widetilde{\Omega}_{n_l}^{n_0}+(\lambda \widetilde{\Omega}_{n_l}^{n_r}+\ldots )\widetilde{\Omega}_{n_r}^{n_0}\bigr)\Bigr)\label{eq: series Pm(s)},
 \end{align}
 where we have used 
 \begin{align}
     \pazocal{L}[\mathrm{e}^{-\lambda t} J_{|m-n'|}^2(\gamma t)]=\frac{1}{\pi\gamma}Q_{|m-n'|-\frac{1}{2}}\left(\theta'\right)\label{eq: laplace_trans of Bessel}
 \end{align} 
 that follows from Eq.~\eqref{eq: Bessel integral}. We have also used Eq.~\eqref{eq:first-reset} to imply that \begin{align}
     \pazocal{L}[ \mathrm{e}^{-\lambda t} \Omega_k^j(\gamma t)]= \widetilde{\Omega}_k^j\left(\theta'\right),\label{eq: Q_tilde(lambda+s)}
 \end{align} 
where we have defined 
 \begin{align}
 \theta'\equiv 1+\frac{(\lambda+s)^2}{2 \gamma^2}.\label{eq: theta_prime}
 \end{align}
 In Eq.~\eqref{eq: series Pm(s)}, we have used $\widetilde{\Omega}_k^j\equiv \widetilde{\Omega}_k^j\left(\theta'\right)$ and $Q\equiv Q\left(\theta'\right)$, wherein we have suppressed the argument $\theta'$. Next, consider Eq.~\eqref{eq:norm-Q}: multiplying both sides of the equation by $\mathrm{e}^{-(s+\lambda)t}$, integrating over $t$ from $0$ to $\infty$ and then using Eq.~\eqref{eq: Q_tilde(lambda+s)}, one gets
 \begin{align}
\widetilde{\Omega}_{n_l}^{n_0}\left(\theta'\right)+\widetilde{\Omega}_{n_r}^{n_0}\left(\theta'\right)=\frac{1}{\lambda+s} ~\forall~n_0.
\label{eq: tot prob in s space}
\end{align}

Now, $\Omega_{k}^{j}(\gamma t)$ being a probability is smaller than unity, and so we have 
\begin{align}
    \lambda \mathrm{e}^{-\lambda t}\Omega_{k}^{j}(\gamma t)<\lambda \mathrm{e}^{-\lambda t}.
\end{align}
It then follows that 
we have
\begin{align}
    &\pazocal{L}[  \lambda \mathrm{e}^{-\lambda t} \Omega_k^j(\gamma t)] < \pazocal{L}[ \lambda \mathrm{e}^{-\lambda t} ]= \frac{\lambda}{\lambda +s}<1,
\end{align}
which implies on using Eq.~\eqref{eq: Q_tilde(lambda+s)} that $ \lambda\widetilde{\Omega}_k^j\left(\theta'\right)<1$.\\
Let us define a matrix $\widetilde{\Omega}$ as 
\begin{align}
\widetilde{\Omega}\equiv    \begin{bmatrix}
\widetilde{\Omega}_{n_r}^{n_r} & \widetilde{\Omega}_{ n_l}^{n_r} \\
\widetilde{\Omega}_{n_r}^{n_l} & \widetilde{\Omega}_{n_l}^{n_l} \\
\end{bmatrix},
\label{eq:Q-matrix}
\end{align}
where the $(jk)$-th element of the matrix $\widetilde{\Omega}$ is given by
\begin{align}
[\widetilde{\Omega}]_{jk}=\widetilde{\Omega}_k^j.
\end{align}
Using the matrix~\eqref{eq:Q-matrix}, Eq.~\eqref{eq: series Pm(s)} may now be written as
 \begin{widetext}
   \begin{align}
\widetilde{\overline{P}}_m(s)&=\frac{1}{\pi \gamma} \Biggl(Q_{|m-n_0|-\frac{1}{2}}+\lambda~Q_{|m-n_r|-\frac{1}{2}}\Bigl(\sum_{r=0}^{\infty}(\lambda)^r ( \widetilde{\Omega}^r)_{n_r}^{n_r} \widetilde{\Omega}_{n_r}^{ n_0} +\sum_{r=1}^{\infty} (\lambda)^r( \widetilde{\Omega}^r)_{n_r}^{n_l}\widetilde{\Omega}_{n_l}^{ n_0}\Bigr)
   \nonumber\\
   &+\lambda~Q_{|m-n_l|-\frac{1}{2}}\Bigl(\sum_{r=0}^{\infty} (\lambda)^r( \widetilde{\Omega}^r)_{n_l}^{n_l}\widetilde{\Omega}_{n_l}^{ n_0}+\sum_{r=1}^{\infty} (\lambda)^r( \widetilde{\Omega}^r)_{n_l}^{n_r} \widetilde{\Omega}_{n_r}^{n_0}\Bigr) \Biggr)\nonumber \\
&= \frac{1}{\pi \gamma}\Biggl(Q_{|m-n_0|-\frac{1}{2}}+\lambda~Q_{|m-n_r|-\frac{1}{2}}\Bigl(\left({{(I-\lambda\widetilde{\Omega}})}^{-1}\right)_{ n_r n_r} \widetilde{\Omega}_{ n_r}^{ n_0}+\left({((\lambda\widetilde{\Omega})^{-1}-I)}^{-1}\right)_{n_l n_r}\widetilde{\Omega}_{n_l}^{ n_0}\Bigr)\nonumber \\
&+\lambda~Q_{|m-n_l|-\frac{1}{2}}\Bigl(\left({{(I-\lambda\widetilde{\Omega}})}^{-1}\right)_{ n_l n_l} \widetilde{\Omega}_{ n_l}^{ n_0}+\left({((\lambda\widetilde{\Omega})^{-1}-I)}^{-1}\right)_{n_r n_l}\widetilde{\Omega}_{n_r}^{ n_0}\Bigr)\Biggr),\label{eq: Pm(lambda+s) 2nd}
\end{align}
\end{widetext}
where the respective matrix elements of the matrices ${{(I-\lambda\widetilde{\Omega}})}^{-1}$ and ${((\lambda\widetilde{\Omega})^{-1}-I)}^{-1}$ are given by
\begin{align}
    &\left({{(I-\lambda\widetilde{\Omega}})}^{-1}\right)_{n_r n_r}\nonumber\\&= \frac{1-\lambda\widetilde{\Omega}_{n_l}^{n_l}}{(1-\lambda\widetilde{\Omega}_{n_r}^{ n_r})(1-\lambda\widetilde{\Omega}_{n_l}^{n_l})-\lambda^2\widetilde{\Omega}_{n_r}^{n_l}\widetilde{\Omega}_{n_l}^{n_r}},
    \label{mat elements: Laplace domain1}
    \end{align}
    
    \begin{align}
        &\left({{(I-\lambda\widetilde{\Omega}})}^{-1}\right)_{n_l n_l}\nonumber\\
        &= \frac{1-\lambda\widetilde{\Omega}_{n_r}^{n_r}}{(1-\lambda\widetilde{\Omega}_{n_r}^{n_r})(1-\lambda\widetilde{\Omega}_{n_l}^{ n_l})-\lambda^2\widetilde{\Omega}_{n_r}^{ n_l}\widetilde{\Omega}_{n_l}^{ n_r}},
        \label{mat elements: Laplace domain2}
    \end{align}
    \begin{align}
        &\left({((\lambda\widetilde{\Omega})^{-1}-I)}^{-1}\right)_{n_l n_r}\nonumber \\
        &= \frac{\lambda\widetilde{\Omega}_{n_r}^{ n_l}}{(1-\lambda\widetilde{\Omega}_{n_r}^{ n_r})(1-\lambda\widetilde{\Omega}_{n_l}^{ n_l})-\lambda^2\widetilde{\Omega}_{n_r}^{n_l}\widetilde{\Omega}_{n_l}^{ n_r}},
        \label{mat elements: Laplace domain3}
    \end{align}       
   \begin{align}
    &\left({((\lambda\widetilde{\Omega})^{-1}-I)}^{-1}\right)_{n_r n_l}\nonumber \\
    &= \frac{\lambda\widetilde{\Omega}_{n_l}^{ n_r}}{(1-\lambda\widetilde{\Omega}_{n_l}^{n_r})(1-\lambda\widetilde{\Omega}_{n_l}^{n_l})-\lambda^2\widetilde{\Omega}_{n_r}^{n_l}\widetilde{\Omega}_{n_l}^{
    n_r}}. \label{mat elements: Laplace domain4}
\end{align}

Using the well-known final value theorem, $\lim_{t\to \infty}f(t)=\lim_{s\to 0}sF(s);~F(s)\equiv \pazocal{L}(f(t))$, yields the stationary-state value from Eq.~\eqref{eq: Pm(lambda+s) 2nd} as
\begin{align}
   \overline{P}_m^{\mathrm{st}}\equiv \lim_{t \to \infty} \overline{P}_m(t)= \lim_{s \to 0} s~\widetilde{\overline{P}}_m(s).
\end{align}
To proceed, let us consider evaluating the matrix element $\left({{(I-\lambda\widetilde{\Omega}})}^{-1}\right)_{n_r n_r}$, which on using Eqs.~\eqref{mat elements: Laplace domain1} together with Eq.~\eqref{eq: tot prob in s space} with $n_0=n_r$ and also with $n_0=n_l$, reads as
\begin{align}
    \left({{(I-\lambda\widetilde{\Omega}})}^{-1}\right)_{n_r n_r}&= \frac{(s+\lambda)^2 (1- \lambda \widetilde{\Omega}_{n_l}^{n_l})}{s(s+2 \lambda -\lambda(s+\lambda)(\widetilde{\Omega}_{n_l}^{n_l}+\widetilde{\Omega}_{n_r}^{n_r}))},
\end{align}
so that 
\begin{align}
     &\lim_{s \to 0}~s\left({{(I-\lambda\widetilde{\Omega}})}^{-1}\right)_{n_r n_r} Q_{|m-n_r|-\frac{1}{2}}\nonumber\\
     &=\frac{\lambda (1-\lambda\widetilde{\Omega}_{n_l}^{n_l}(\theta))}{2-\lambda(\widetilde{\Omega}_{n_l}^{n_l}(\theta)+\widetilde{\Omega}_{n_r}^{n_r}(\theta))} Q_{|m-n_r|-\frac{1}{2}}(\theta),
\end{align}
where note that  $Q_{|m-n_r|-\frac{1}{2}}$ and $\widetilde{\Omega}_k^j$'s are functions of $\theta'$ defined in Eq.~\eqref{eq: theta_prime}, and so after taking the limit $s\to0$ become functions of $\theta$ defined in Eq.~\eqref{eq: def_theta}.

In a similar manner, the matrix elements in Eqs.~\eqref{mat elements: Laplace domain2},~\eqref{mat elements: Laplace domain3},~\eqref{mat elements: Laplace domain4} may be evaluated. Equation~\eqref{eq: Pm(lambda+s) 2nd} then yields the stationary-state probability of finding the TBM particle on site $m$, given an arbitrary initial location $n_0$ and arbitrary reset sites $n_r$ and $n_l$, as
\begin{widetext}
    \begin{align}      \overline{P}_m^{\mathrm{st}}&= \frac{\lambda^2 }{\pi \gamma}\left(\frac{(1-\lambda\widetilde{\Omega}_{n_l}^{n_l}(\theta))(\widetilde{\Omega}_{n_r}^{n_0}(\theta)+\widetilde{\Omega}_{n_l}^{n_0}(\theta))}{2-\lambda(\widetilde{\Omega}_{n_l}^{n_l}(\theta)+\widetilde{\Omega}_{n_r}^{n_r}(\theta))} Q_{|m-n_r|-\frac{1}{2}}(\theta) + \frac{(1-\lambda\widetilde{\Omega}_{n_r}^{n_r}(\theta))(\widetilde{\Omega}_{n_r}^{n_0}(\theta)+\widetilde{\Omega}_{n_l}^{n_0}(\theta))}{2-\lambda(\widetilde{\Omega}_{n_l}^{n_l}(\theta)+\widetilde{\Omega}_{n_r}^{n_r}(\theta))}Q_{|m-n_l|-\frac{1}{2}}(\theta)\right),
\end{align}
\end{widetext}
where we have used the fact that the first term in Eq.~\eqref{eq: Pm(lambda+s) 2nd} vanishes upon taking the limit $s\to 0$. On using Eq.~\eqref{eq: tot prob in s space} with $s=0$ for $n_0$, $n_0=n_r$ and $n_0=n_l$, we finally get
\begin{align}      \overline{P}_m^{\mathrm{st}}&= \frac{\lambda}{\pi\gamma}\Biggl(\frac{\widetilde{\Omega}_{n_r}^{n_l}(\theta)}{\widetilde{\Omega}_{n_r}^{n_l}(\theta)+\widetilde{\Omega}_{n_l}^{n_r}(\theta)}Q_{|m-n_r|-\frac{1}{2}}(\theta) \nonumber\\
& + \frac{\widetilde{\Omega}_{n_l}^{n_r}(\theta)}{\widetilde{\Omega}_{n_r}^{n_l}(\theta)+\widetilde{\Omega}_{n_l}^{n_r}(\theta)}Q_{|m-n_l|-\frac{1}{2}}(\theta)\Biggr),\label{eq: Pmst general n0}
\end{align}
where $\widetilde{\Omega}_{\mu}^{j}(\theta)$ (with $\mu=n_r$ or $n_l$) is given from Eqs.~\eqref{eq:first-reset} and \eqref{eq: laplace_trans of Bessel} as
\begin{align}
    \widetilde{\Omega}_{n_r}^{j}(\theta)=\frac{1}{\pi\gamma}\sum_{k=1}^{\infty} Q_{|k-j|-\frac{1}{2}}(\theta) +\frac{1}{2\pi\gamma}Q_{|j|-\frac{1}{2}}(\theta),\nonumber\\
    \\
     \widetilde{\Omega}_{n_l}^{j}(\theta)=\frac{1}{\pi \gamma}\sum_{k=-1}^{-\infty} Q_{|k-j|-\frac{1}{2}}(\theta) +\frac{1}{2\pi\gamma}Q_{|j|-\frac{1}{2}}(\theta).\nonumber
\end{align}
Note that, as desired, the stationary-state probability in Eq.~\eqref{eq: Pmst general n0} does not depend on the choice of the initial site $n_0$.~However, Eq.~\eqref{alpha series time domain} along with Eq.~\eqref{eq: general Pm alpha} imply that the time-dependent probability $\overline{P}_m(t)$ quite expectedly does depend on $n_0$. We remark that on putting $n_l=-n_r$ in Eq.~\eqref{eq: Pmst general n0}, together with the associated equality $\widetilde{\Omega}_{n_l}^{n_r}(\theta)=\widetilde{\Omega}_{n_r}^{n_l}(\theta)$ (which follows from the fact that we have $\Omega_{n_l}^{n_r}(t)=\Omega_{n_r}^{n_l}(t)$), correctly recovers  Eq.~\eqref{Eq: Stationary state probability}. A result similar in structure to Eq.~\eqref{eq: Pmst general n0} has been obtained in Ref.~\cite{PhysrevB.104.L180302} in the context of the transverse-field quantum
Ising chain.

We may wish to obtain the first two moments of the probability given by Eq.~\eqref{eq: Pmst general n0}. The  mean displacement from the initial location $n_0$ in the stationary state, given by $\overline{[m - n_0]}^{\mathrm{st}}=\sum_{m=-\infty}^\infty (m-n_0)\overline{P}_m^{\mathrm{st}}$, may be calculated from Eq.~\eqref{eq: Pmst general n0} on using Eq.~\eqref{eq: Bessel integral} and  $\sum_{m=-\infty}^\infty m J^2_m(t)=0$, as
\begin{align}
   \overline{[m - n_0]}^\mathrm{st}&= \frac{\widetilde{\Omega}_{n_r}^{n_l}(\theta)}{\widetilde{\Omega}_{n_r}^{n_l}(\theta)+\widetilde{\Omega}_{n_l}^{n_r}(\theta)}(n_r-n_0) \nonumber\\
& + \frac{\widetilde{\Omega}_{n_l}^{n_r}(\theta)}{\widetilde{\Omega}_{n_r}^{n_l}(\theta)+\widetilde{\Omega}_{n_l}^{n_r}(\theta)}(n_l-n_0).\label{eq: Mean_st general n0}
\end{align}
The stationary-state MSD is readily calculated from Eqs.~\eqref{eq: Pmst general n0} by using Eq.~\eqref{eq: MSD_TBM_noreset} and \eqref{eq: laplace_trans of Bessel}, to obtain
\begin{align}
    \overline{S}^\mathrm{st}&= \frac{\widetilde{\Omega}_{n_r}^{n_l}(\theta)}{\widetilde{\Omega}_{n_r}^{n_l}(\theta)+\widetilde{\Omega}_{n_l}^{n_r}(\theta)}\left(n_r^2-2 n_0 n_r +n_0^2+ \frac{\gamma^2}{\lambda^2}\right)  \nonumber\\
& + \frac{\widetilde{\Omega}_{n_l}^{n_r}(\theta)}{\widetilde{\Omega}_{n_r}^{n_l}(\theta)+\widetilde{\Omega}_{n_l}^{n_r}(\theta)}\left(n_l^2-2 n_0 n_l +n_0^2+ \frac{\gamma^2}{\lambda^2}\right).\label{eq: MSD_st general n0}
\end{align}
 Equations~\eqref{eq: Mean_st general n0} and~\eqref{eq: MSD_st general n0} recover Eqs.~\eqref{eq:mean-center-stationary} and~ \eqref{eq:MSD-center-stationary}, respectively, upon using $n_l=-n_r$, $n_0=0$, and $\widetilde{\Omega}_{n_l}^{n_r}(\theta)=\widetilde{\Omega}_{n_r}^{n_l}(\theta)$.

\subsection{Effects of an additional periodic external forcing}
We now discuss the case of a time-dependent Hamiltonian subject to our protocol of conditional resetting. To achieve our goal, we consider the case of the TBM with the particle being charged (with charge $q$) and in the presence of a periodic electric field $E(t)=E_0 \cos(\omega_0 t)$. The force acting on the particle being $F(x)= q E(t)$, corresponding to a potential $-qE(t)x;~x=\sum_{n=-\infty}^\infty n|n\rangle \langle n|$, the TBM Hamiltonian is changed from Eq.~\eqref{eq:H} to a time-dependent Hamiltonian 
\begin{align}
    H(t)&=-\dfrac{\gamma}{2}\sum_{n=-\infty}^{\infty}\left(|n\rangle \langle n+1 | + |n+1\rangle \langle n | \right)\nonumber\\
    &+ F_0 \cos(\omega_0 t)\sum_{n=-\infty}^{\infty}n|n\rangle \langle n |,\label{eq:time_dep_H}
\end{align}
with $F_0\equiv -q E_0$. In addition to the previously-defined operators $K$ and $K^{\dagger}$, see Eq.~\eqref{eq:KKdagger}, let us introduce a new operator $N \equiv \sum_{n=-\infty}^{\infty}n |n\rangle \langle n |$, which is diagonal in the Wannier states $|n\rangle$. Equation~\eqref{eq:time_dep_H} then rewrites as 
\begin{align}
H(t)=H+H_0(t);~H_0(t) \equiv F_0 \cos(\omega_0t)N,
\end{align}
with the time-independent part of the Hamiltonian denoted by $H$ and given by Eq.~\eqref{eq:H-KKdagger}. Since we have $H(t)=H(t+\mathcal{T})$, with $\mathcal{T}$ being the time period of the oscillatory forcing field, $H(t)$ is a Floquet-type Hamiltonian~\cite{doi:10.1080/00018732.2015.1055918}. Now, we have $[K,N] = K$ and $[K^\dagger,N] = -K^\dagger$, and hence, the two terms in the above Hamiltonian do not commute with each other. Moreover, $H(t)$ for two different times do not commute: $[H(t),H(t')]=(\gamma/2)F_0(\cos(\omega_0t)-\cos(\omega_0t'))(K-K^\dagger)$, thus necessitating the use of time ordering in the formalism developed in Section~\ref{sec:general-set-up}.

We may now ask the same question as in the case of zero forcing dealt with in Section~\ref{sec: Application of the protocol to Tight Binding Model}: Given that the particle at the initial instant $t=0$ was on site $n_0$, what is the probability $P_m(t)$ of finding it on an arbitrary site $m$ at a later time $t>0$ ? The detailed calculation for finding $P_m(t)$ is presented in Ref~\cite{Debraj:2022-2}, which we briefly summarize here. We have, by definition, that $P_m(t)=\langle m|\rho(t)|m\rangle$, with the density operator $\rho(t)$ evolving while starting from its initial form $\rho(0)=|n_0\rangle \langle n_0|$ as
$\rho(t)=\mathrm{e}_+^{-\mathrm{i} \int_0^t \mathrm{d}t'~H(t')}\,\rho(0) \, \mathrm{e}_-^{\mathrm{i} \int_0^t \mathrm{d}t'~H(t')}$. Invoking the interaction picture of time evolution in quantum mechanics, one transforms $\rho(t)$ to $\widetilde{\rho}(t)$, as 
\begin{align}
\widetilde{\rho}(t)\equiv \mathrm{e}^{\mathrm{i} \int_0^t \mathrm{d}t'~H_0(t')}\rho(t) \, \mathrm{e}^{-\mathrm{i} \int_0^t \mathrm{d}t'~H_0(t')}, 
\label{eq:rho-transformation}
\end{align}
 where the commutation of $H_0(t)$ at two different times implies that the exponential factors can be written without time ordering.  Since $H_0(t)$ is diagonal in the Wannier states,  we have $P_m(t)=
\langle m|\rho(t)|m\rangle=\langle m|\widetilde{\rho}(t)|m\rangle$. One finds that the transformed density operator $\widetilde{\rho}(t)$ evolves according to the Liouville equation~\cite{Debraj:2022-2}
\begin{align}
\frac{\partial \widetilde{\rho}(t)}{\partial t}=-\mathrm{i}[H_\mathrm{I}(t),\widetilde{\rho}(t)],
\label{eq:rhotilde}
\end{align}
with $ H_\mathrm{I}(t)$ being the Hamiltonian $H$ in the interaction picture, defined by
\begin{align}
    H_\mathrm{I}(t)\equiv  \mathrm{e}^{\mathrm{i} \int_0^t \mathrm{d}t'~H_0(t')} H \, \mathrm{e}^{-\mathrm{i} \int_0^t \mathrm{d}t'~H_0(t')}.\label{def: H interaction pic}
\end{align}
Equation~\eqref{def: H interaction pic} when simplified using the well-known Baker–Campbell–Hausdorff formula for $2$ operators $A$ and $B$,~given by $\mathrm{e}^A B \mathrm{e}^{-A} =B+[A,B]+[A,[A,B]]/2 + \ldots$, yields
\begin{align}
H_\mathrm{I}(t)&= -\frac{\gamma}{2}\left(K \mathrm{e}^{-\mathrm{i} \eta(t)}+K^\dagger \mathrm{e}^{\mathrm{i} \eta(t)}\right),
\label{eq:HTBLt}
\end{align}
where we have defined $\eta(t) \equiv (F_0/\omega_0) \sin (\omega_0 t)$, and have
 $[H_\mathrm{I}(t),H_\mathrm{I}(t')]=0$. Equation~\eqref{eq:rhotilde} has the solution
\begin{align}
\widetilde{\rho}(t)=\mathrm{e}^{-\mathrm{i} \int_0^t \mathrm{d}t'~H_\mathrm{I}(t')}\rho(0) \, \mathrm{e}^{\mathrm{i} \int_0^t \mathrm{d}t'~H_\mathrm{I}(t')},
\label{eq:rhotilde-solution}
\end{align}
so that
\begin{align}
P_m(t)=\langle m|\mathrm{e}^{-\mathrm{i} \int_0^t \mathrm{d}t'~H_\mathrm{I}(t')}|n_0\rangle \langle n_0|\mathrm{e}^{\mathrm{i} \int_0^t \mathrm{d}t'~H_\mathrm{I}(t')}|m\rangle.
\label{eq:Pmt-definition}
\end{align}
Next, using Eq.~\eqref{eq:HTBLt}, it may be shown that~\cite{Debraj:2022-2}
\begin{align}
\mathrm{e}^{\mathrm{i} \int_0^t \mathrm{d}t'~H_\mathrm{I}(t')}= \mathrm{e}^{-\mathrm{i} (\gamma/2)[Kw^\star(t)+K^\dagger w(t)]},
\label{eq:expHTBLt}
\end{align}
with
\begin{align}
&w(t) \equiv \int_0^t \mathrm{d}t'~ \mathrm{e}^{\mathrm{i} \eta(t')}=\frac{1}{\omega_0}\int_0^{\omega t}\mathrm{d}\tau~\sum_{p=-\infty}^\infty J_p(F_0/\omega_0) \mathrm{e}^{\mathrm{i} p \tau}. \label{eq: tau integral}
\end{align}
For $t=n\mathcal{T}$,  so that,  $\omega_0 t=2\pi n$,  with $n$ being an integer,  the integral in Eq.~\eqref{eq: tau integral} vanishes unless $p=0$, and one obtains
\begin{align}
w(n\mathcal{T})=tJ_0(F_0/\omega_0).
\label{eq:wmT}
\end{align}
From Eq.~\eqref{eq:Pmt-definition}, we get on transforming to the Bloch states and using Eqs.~\eqref{eq:expHTBLt},~\eqref{eq:K-eigenvalue},~\eqref{free_prop} and the result $\langle k'|k\rangle=\delta(k-k')$ that~\cite{Debraj:2022-2}
\begin{align}
P_m(t)=J^2_{|m-n_0|}(\gamma |w(t)|).
\label{eq:Pmt-final}
\end{align}
It is easily checked that in the absence of forcing, one has $w(t)=t$, and the above equation reduces to the result~\eqref{free_prop}.

In the above backdrop, let us study the effect of conditional stochastic resetting, by applying the formalism developed in Section~\ref{sec:general-set-up}. Comparing Eqs.~\eqref{eq:Pmt-final} and ~\eqref{free_prop}, one may conclude that analytical results in the presence of forcing may be obtained from those in its absence through the mere substitution $t \to |w(t)|$ as the argument of all occurrences of the Bessel function in the latter analysis.
 With this proviso, we present now an illustration of explicit results so obtained in the presence of forcing, by considering the specific scenario of the initial location of the TBM particle being $n_0=0$, and moreover, with equidistant reset locations around the origin, namely, $n_l=-n_r$. Using Eq.~\eqref{eq: P_tot(t)_assumption}, we thus obtain the site-occupation probability, in the presence of forcing and conditional resetting, for the TBM particle to be found on site $m$ at time $t$ as
 \begin{align}
      &\overline{P}_m (t)
     = \mathrm{e}^{-\lambda t} J_{|m-n_0|}^2(\gamma~|w(t)|) \nonumber \\&+ \dfrac{\lambda}{2}\hspace{-0.1 cm}\int_0^t \hspace{-0.2 cm}\mathrm{d}t_1 \mathrm{e}^{-\lambda t_1}( J^2_{|m-n_r|}(\gamma ~|w(t_1)|)+ J^2_{|m+n_r|}(\gamma ~|w (t_1)|)).\label{eq: P_tot(t)_TDH_midstart}
\end{align}
The corresponding stationary-state site-occupation probability is obtained as
 \begin{align}
&\overline{P}_m^\mathrm{st}= \dfrac{\lambda}{2}\:\int_0^\infty \mathrm{d}t_1 \:\mathrm{e}^{-\lambda t_1}\nonumber \\
&\times ( J^2_{|m-n_r|}(\gamma ~|w(t_1)|)+ J^2_{|m+n_r|}(\gamma ~|w (t_1)|)),\label{eq: P_tot(t)_TDH_midstart-steady-state}
\end{align}
which may be used to obtain numerically the MSD in the stationary state. 

In the following, we report on physical
implications of our derived analytical results for the site-occupation probability.  
\subsection{Results and discussions}
\label{sec:numerics}

This subsection is devoted to a discussion of physical implications of our derived analytical results for the site-occupation probability and other quantities and their validation via numerical implementation of the dynamics. We start with a discussion of our scheme of numerical implementation of the TBM dynamics subject to conditional resetting at random times. We follow this up with a discussion of our numerical results obtained within such a scheme. We consider a one-dimensional open lattice with $(2N+1)$ sites labeled $\{-N,-N+1,\ldots,0,\ldots,N-1,N\}$. The density operator and the Hamiltonian operator are both given in terms of $(2N+1) \times (2N+1)$ matrices, while the interaction superoperator $T(\tau)$ is a matrix of dimension $(2N+1)^2 \times (2N+1)^2$. In order to implement a particular realization of the evolution of the system over a total duration $t$, we first enumerate the time intervals $\tau$ between successive resets, which is done by sampling independently these numbers from the exponential distribution~\eqref{eq: exp_dist} (or, the power-law distribution~\eqref{eq: power_law_dist}, see later) by using standard techniques.  Suppose in a given realization there happens to be ${\pazocal{N}}$ number of resets, so that the $\tau$'s satisfy $\sum_{\alpha=1}^{\pazocal{N}} \tau_\alpha < t$ and that $\tau_{{\pazocal{N}}+1} = t-\sum_{\alpha=1}^{\pazocal{N}} \tau_\alpha$.  Here, the quantity $\tau_\alpha$ denotes the time interval between the $\alpha$-th and the $(\alpha-1)$-th reset. Once the $\tau_\alpha$'s are known, the dynamical evolution with the TBM particle starting initially from the location $n_0$ may be implemented as follows: The initial density operator $\rho(0)=|n_0\rangle \langle n_0|$, which has only its $(n_0,n_0)$-th element nonzero and equal to unity, is allowed to evolve unitarily in time for time duration $\tau_1$ to obtain the evolved density operator as $\rho(\tau_1)=\exp_+(-\mathrm{i}\int_0^{\tau_1}\mathrm{d}t'~H(t'))\rho(0)\exp_-(\mathrm{i}\int_0^{\tau_1}\mathrm{d}t'~H(t'))$. Depending on whether we want to study the TBM in the absence or in presence of the periodic forcing, the Hamiltonian is either the time-independent Hamiltonian~\eqref{eq:H} (in which case, no time ordering in performing unitary evolution of the density operator has to be invoked) or the time-dependent Hamiltonian~\eqref{eq:time_dep_H}, respectively. In order to now have a conditional resetting of the system to the two sites $n_r$ and $n_l$, we now operate on the time-evolved density operator by the interaction operator $T(\tau_1)$, whose explicit form is given by Eq.~\eqref{eq: T1(t)}. The resulting matrix is then evolved unitarily for time $\tau_2$. Next, we operate on the matrix obtained following the evolution by the interaction operator $T(\tau_2)$ given in Eq.~\eqref{eq: T2(t)}, and so on.  After an ${\pazocal{N}}$ number of resets have been performed on the system, the final step of evolution involves unitary evolution for time duration $\tau_{{\pazocal{N}}+1}$. The obtained final density matrix qualifies as the density matrix in the given realization of the dynamics of the TBM subject to conditional resetting, and its $(m,m)$-th element yields the probability, in the given dynamical realization, for the TBM particle to be found on site $m$ at time $t$.  The whole process of starting with $\rho(0)$ and letting it undergo unitary evolution interspersed with instantaneous conditional resets is repeated for several realizations of the random times $\tau_\alpha$ in order to obtain the averaged site-occupation probability $\overline{P}_m(t)$ for the particle to be on site $m$ at time $t$. 

\begin{figure}[H]
\includegraphics[width=1\linewidth]{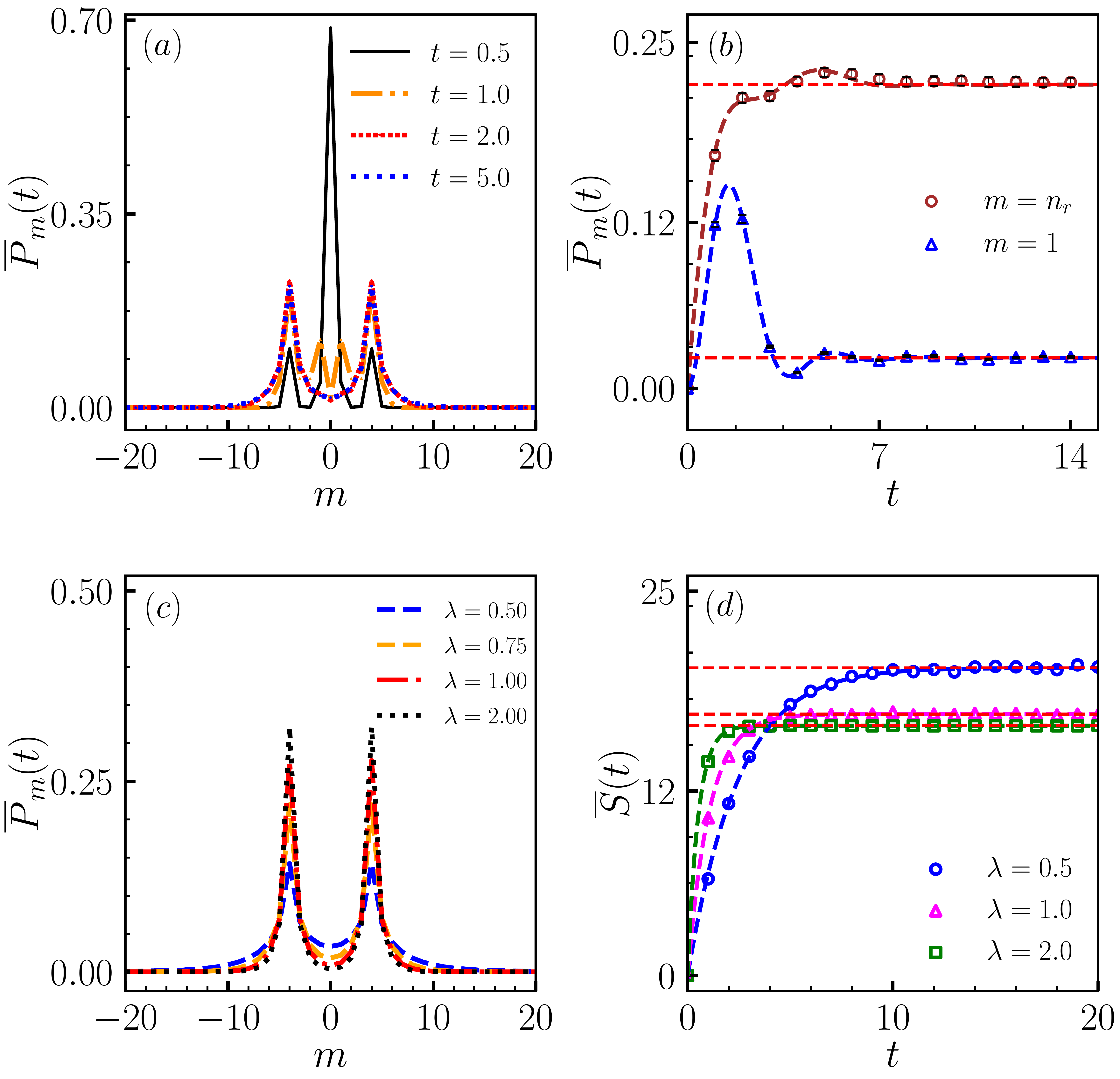}
\caption{For the TBM system~\eqref{eq:H} subject to conditional resetting at exponentially-distributed time intervals, the figure corresponds to the case of the initial location of the TBM particle being at the origin, $n_0=0$, and with resetting to locations symmetrically disposed with respect to the initial location, namely, $n_r=4$ and $n_l=-4$. Thus, the reset sites are equidistant with respect to the initial location, which is taken to be the origin. Panel (a) shows analytical results given by Eq.~\eqref{eq: P_tot(t)_assumption} for the site-occupation probability $\overline{P}_m(t)$ plotted against $m$ at four different times, with $\lambda=0.5$. As time progresses, the system attains a stationary state characterized by localization of the particle around the two reset sites, as evident from the data for $t=2.0$ and $t=5.0$. Panel (b) shows $\overline{P}_m(t)$ versus $t$ for two different $m$'s, one being the reset site $m=n_r$ and another being a non-reset site $m=1$, with $\lambda=0.5$. The analytical results given by Eq.~\eqref{eq: P_tot(t)_assumption} and plotted with lines are compared with numerical results depicted by symbols. The latter results are obtained via numerical implementation of the dynamics on a lattice of $41$ sites and involve averaging over $5\times10^3$ dynamical realizations (see Section~\ref{sec:numerics} for details on our scheme of numerical implementation). The horizontal red dashed lines denote analytical stationary-state probability $\overline{P}_m^\mathrm{st}$ given by Eq.~\eqref{Eq: Stationary state probability}. Panel (c) shows analytical results for $\overline{P}_m(t)$ given by Eq.~\eqref{eq: P_tot(t)_assumption} and plotted as a function of $m$ for four different values of $\lambda$ and at large times (namely, at $t=20$), exhibiting enhanced localization with increased $\lambda$. Panel (d) shows analytical (lines) and numerical (symbols) results for the MSD $\overline{S}(t)$, Eq.~\eqref{eq: MSD_reset_with_assumptions}, as a function of time $t$ and for three different values of $\lambda$, implying saturation of the MSD due to localization at long times. The horizontal red dashed lines denote analytical stationary-state MSD given by Eq.~\eqref{eq:MSD-center-stationary}. Consistent with the results in panel (c), one observes that the saturation value of the MSD  decreases with increased $\lambda$. In all cases, we have taken $\gamma=1.0$.}\label{fig: 4}
\end{figure}

We now move on to discuss our results. Figure~\ref{fig: 4} corresponds to the case of the TBM system~\eqref{eq:H} subject to conditional resetting at exponentially-distributed random time intervals. The initial location of the particle is at the origin, while the reset sites $n_r$ and $n_l$ are equidistant from it. Panel (a) shows results for the site-occupation probability $\overline{P}_m(t)$ versus $m$ for different times, short and long, and at a fixed resetting rate $\lambda$. The first thing to note is that in sharp contrast to the bare model results in Fig.~\ref{fig: 2}, the system at long times relaxes to a stationary state characterized by a time-independent site-occupation probability. Moreover, in the latter state, the TBM particle is most likely to be found at either the site $n_r$ or the site $n_l$ with equal probability, while the further one is from the reset sites, the lower is the probability for the particle to be found there (see panel (b)). We thus have an effective localization of the particle around the reset sites. The localization becomes more pronounced with an increase of $\lambda$, as may be seen from the plot in panel (c) that corresponds to the site-occupation probability at long times and for different values of $\lambda$. This localization is induced not by the presence of any boundaries on the lattice that constrain the motion of the particle, but by the act of subjecting the particle to conditional resetting to sites $n_r$ and $n_l$ at random times. Panel (b) demonstrates that the analytical results derived by us for $\overline{P}_m(t)$ match perfectly well with numerical results obtained on the basis of the numerical implementation of the dynamics. Panel (d) is a plot of the MSD, Eq.~\eqref{eq: MSD_reset_with_assumptions}, showing firstly a relaxation to a stationary-state behavior at long times, and secondly, a stationary-state value that decreases with increasing $\lambda$. The latter feature is a manifestation of enhanced localization with increasing $\lambda$, as evident from panel (c). 

\begin{figure}[H]
\includegraphics[width=1\linewidth]{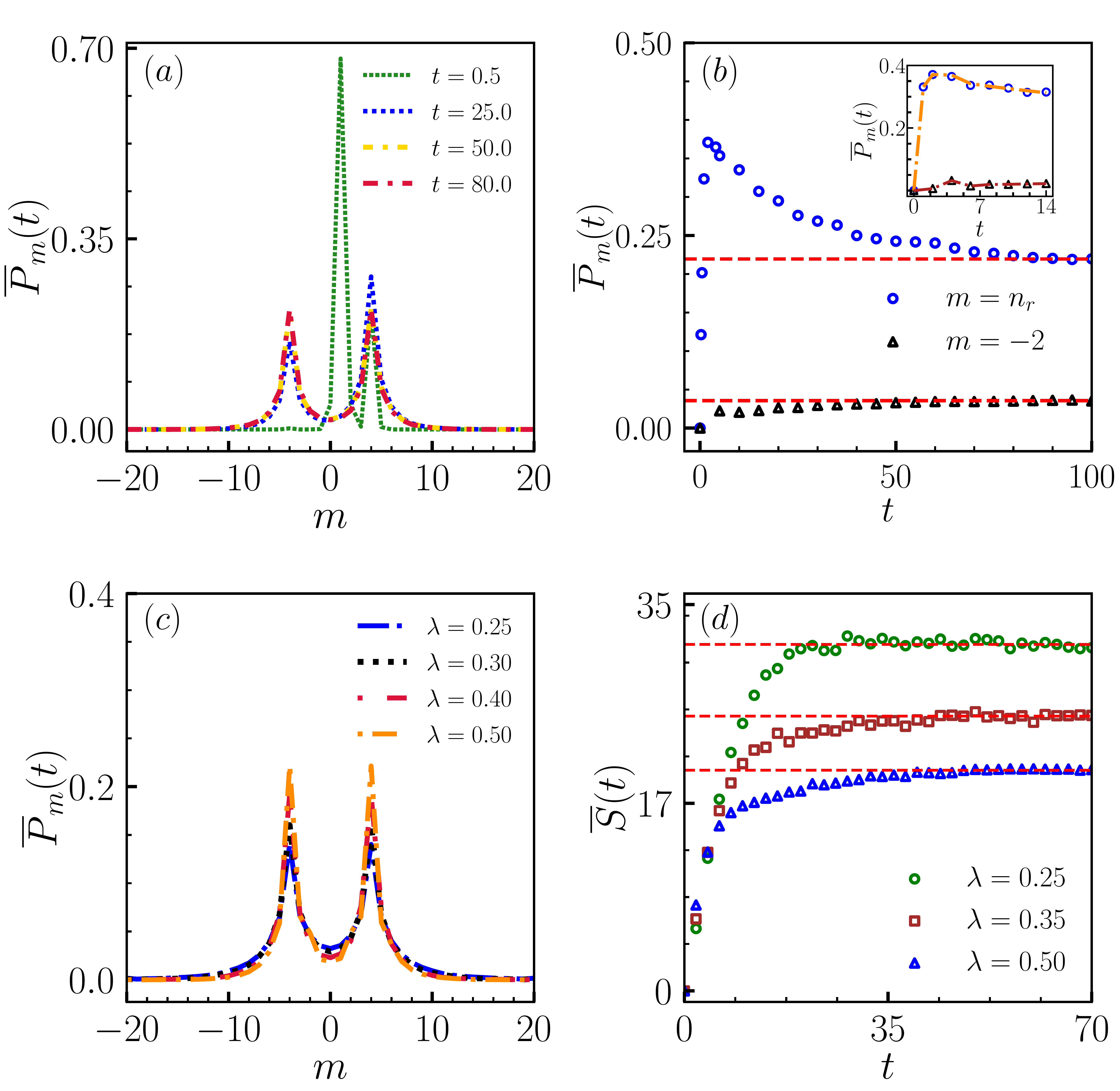}
\caption{For the same TBM system as in Fig.~\ref{fig: 4}, the figure corresponds to the case of the initial location of the TBM particle being $n_0=1$, i.e., not at the origin, and with resetting to locations not symmetrically disposed with respect to the initial location, namely, $n_r=4$ and $n_l=-4$. Thus, the reset sites are not equidistant with respect to the initial location, though they are symmetric with respect to the origin. Panel (a) shows numerical results for the site-occupation probability $\overline{P}_m(t)$ plotted against $m$ at four different times, with $\lambda=0.5$. As time progresses, the system attains a stationary state characterized by localization of the particle around the two reset sites, as evident from the data for $t=50.0$ and $t=80.0$. The numerical results are obtained via numerical implementation of the dynamics on a lattice of $41$ sites and involve averaging over $5\times10^3$ dynamical realizations. Panel (b) shows numerical results for $\overline{P}_m(t)$ versus $t$ for two different $m$'s, one being the reset site $m=n_r$ and another being a non-reset site $m=-2$, with $\lambda$ being $0.5$. The inset shows the comparison between the analytical results given by Eq.~\eqref{alpha series time domain} and plotted with lines and the numerical results depicted by symbols.  The horizontal red dashed lines denote the analytical stationary-state probability $\overline{P}_m^\mathrm{st}$ given by Eq.~\eqref{eq: Pmst general n0}. Panel (c) shows numerical results for $\overline{P}_m(t)$ plotted as a function of $m$ for four different values of $\lambda$ and at large times (namely, at $t=100$), exhibiting enhanced localization with increased $\lambda$. Panel (d) shows numerical (symbols) results for the MSD $\overline{S}(t)$, as a function of time $t$ and for three different values of $\lambda$, implying saturation of the MSD due to localization at long times. The horizontal red dashed lines denote analytical stationary-state MSD given by Eq.~\eqref{eq: MSD_st general n0}. Consistent with the results in panel (c), one observes that the saturation value of the MSD  decreases with increased $\lambda$.}
        \label{fig: 5}
\end{figure}

All of the aforementioned stationary-state features pertaining to reset sites equidistant from the initial location of the TBM particle also hold when the sites are \textit{not} equidistant with respect to the initial location (although they are equidistant with respect to the origin), as illustrated in Fig.~\ref{fig: 5} that corresponds to the initial location being $n_0=1$, while the reset sites are $n_r=4, n_l=-4$. There are differences as well. Since the initial site is now closer to one reset site than the other (namely, closer to $n_r$ than to $n_l$), it is evident that during the initial stage of evolution, the particle is to be found to the right of the origin with a higher probability than to the left (see the data for $t=0.5$ and $t=25.0$ in Fig.~\ref{fig: 5}(a)). Nonetheless, in the stationary state, the particle is equally likely to be found at the two reset sites, similar to the equidistant-reset-site case; this fact is consistent with our expectation that the stationary state is independent of the choice of the initial location of the particle. The fact of initial affinity for the sites to the right of the origin as compared to the left affects the time of relaxation to the stationary state. Indeed, it takes longer to reach the stationary state in the non-equidistant case as compared to the equidistant case (compare the data for $t=2.0$ and $t=5.0$ in panel (a) in Fig.~\ref{fig: 4} and the date for $t=50.0$ and $t=80.0$ in Fig.~\ref{fig: 5}).

\begin{figure}[H]
\includegraphics[width=1\linewidth]{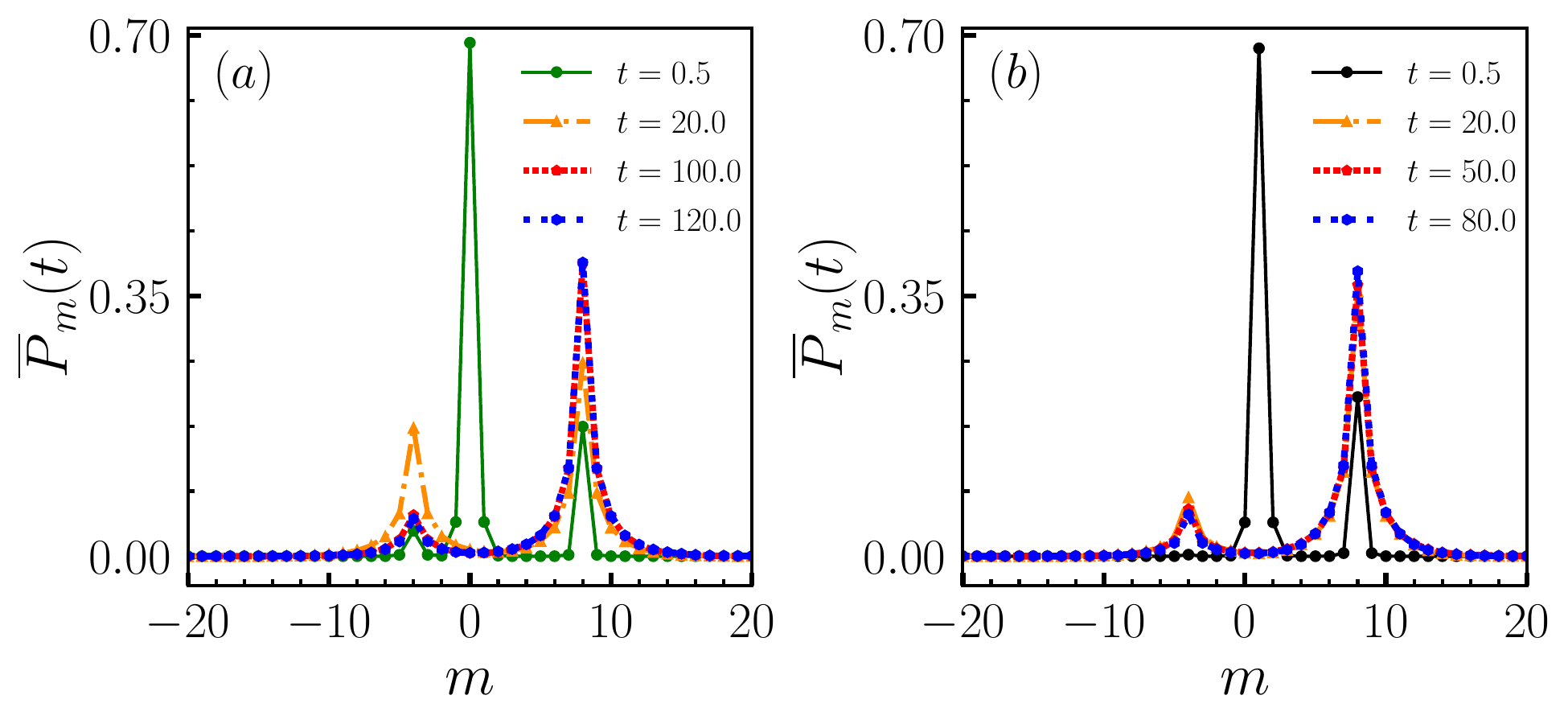}
\caption{For the same TBM system as in Fig.~\ref{fig: 4}, the figure corresponds to the resetting locations not symmetrically disposed with respect to the origin, namely, $n_r=8$ and $n_l=-4$. Thus, the reset sites are not equidistant with respect to the origin. Panel (a) shows numerical results for the site-occupation probability $\overline{P}_m(t)$ plotted against $m$ at four different times, with $\lambda=0.5$ and the initial location at the origin $n_0=0$. Over time, the system attains a stationary state characterized by localization of the particle around the two reset sites, as evident from the data for $t=100.0$ and $t=120.0$. Panel (b) shows numerical results for the site-occupation probability $\overline{P}_m(t)$ plotted against $m$ at four different times, with $\lambda=0.5$ and the initial location at $n_0=1$. Over time, the system reaches a stationary state characterized by the particle localized around the two reset sites, as clearly demonstrated by the data for $t=50.0$ and $t=80.0$. The numerical results are obtained via numerical implementation of the dynamics on a lattice of $41$ sites, and involve averaging over $5\times10^3$ dynamical realizations.}\label{fig: 6}
\end{figure}

Next, we turn our attention to Fig.~\ref{fig: 6}, showing results for two distinct initial locations: $n_0=0$ (origin) in panel (a), and $n_0=1$ in panel (b), and with asymmetrically-disposed reset locations around the origin: $n_r=8$ and $n_l=-4$. Thus,  $n_l$ is closer to the origin than $n_r$, and hence, in contrast to the cases treated in Figs.~\ref{fig: 4} and~\ref{fig: 5}, the reset sites are not equidistant with respect to the origin. We find that at large times, the site-occupation probabilities of the particle at the two reset sites are unequal (see the data for $n_l$ and $n_r$ in panel (a) and panel (b) in Fig.~\ref{fig: 6}), which may be contrasted with equal site-occupation probabilities at two reset sites at large times depicted in Figs.~\ref{fig: 4} and~\ref{fig: 5}. This difference may be understood as follows. The site $n_l$ being closer to the origin, at any given time, the probability for the TBM particle to be found right of the origin while starting from $n_l$ is larger than the same for the TBM particle to be found left of the origin starting from $n_r$. Consequently, the probability to reset to site $n_r$ is larger than the probability to reset to $n_l$. This results in the site-occupation probability being higher to the right of the origin than to the left. Furthermore, that the stationary-state site-occupation probability does not depend on the choice of the initial location is supported by the observation regarding the independence of long-time results with respect to the choice of $n_0$ (as shown in panel (a) for $n_0=0$ and panel (b) for $n_0=1$ of Fig.~\ref{fig: 6}). Both of the last two features are in conformity with Eq.~\eqref{eq: Pmst general n0} in which one has $\widetilde{\Omega}_{n_r}^{n_l}(\theta)>\widetilde{\Omega}_{n_l}^{n_r}(\theta)$. The result of the aforementioned unequal site-occupation probability around the reset sites is quite interesting as it implies that even though there is no bias in the inherent TBM dynamics, an effective bias towards one of the reset sites is nonetheless being induced by the protocol of conditional resetting.

\begin{figure}[H]
\includegraphics[width=1\linewidth]{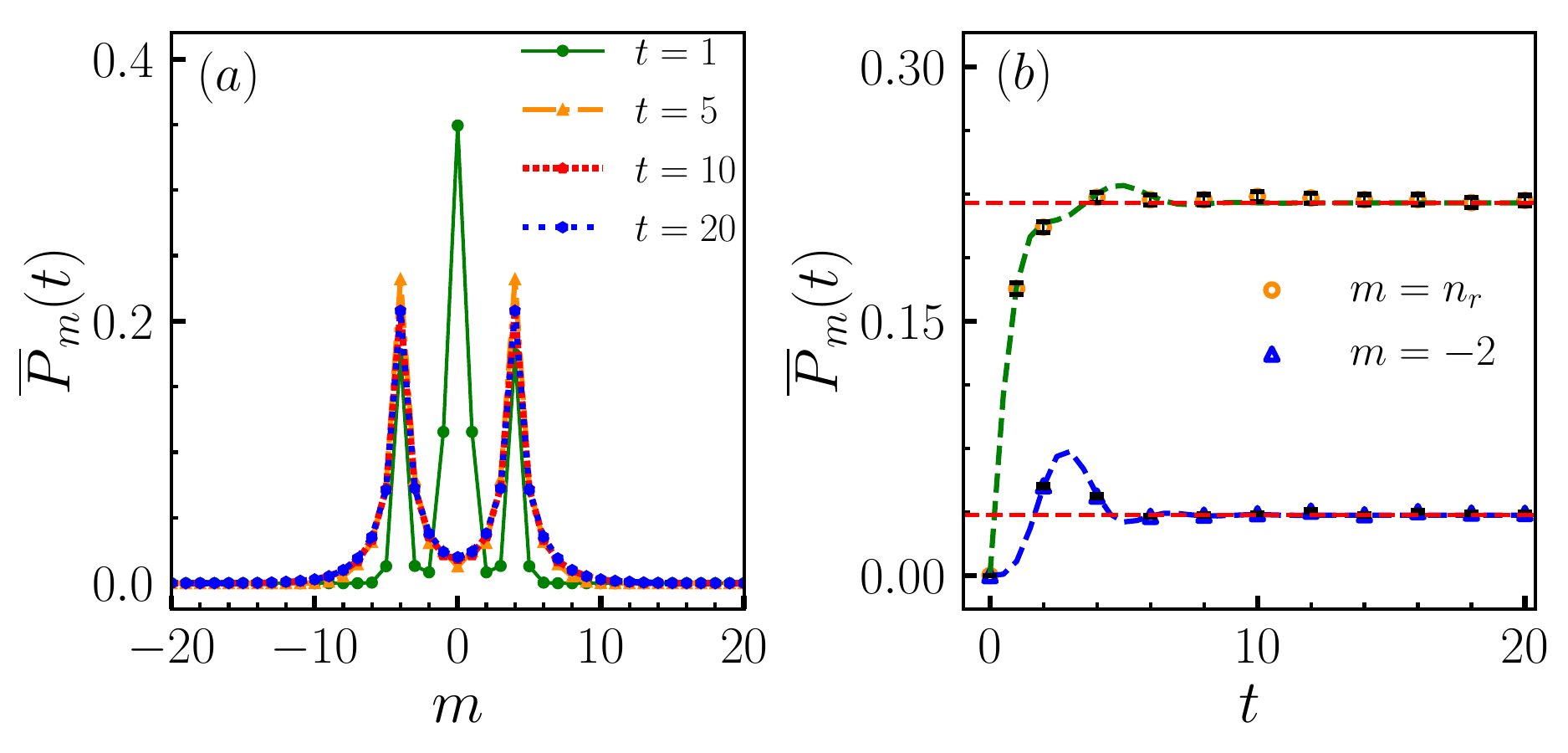}
\caption{For the same TBM lattice as in Fig.~\ref{fig: 4} subject to conditional resetting at exponentially-distributed time intervals but in the presence of a forcing field and described by the time-dependent Hamiltonian given by Eq.~\eqref{eq:time_dep_H}, the figure corresponds to the case of the initial location of the TBM particle being at the origin, $n_0=0$, with resetting to locations equidistant from the initial location, namely, $n_r=4$ and $n_l=-4$. Panel (a) shows analytical results given by Eq.~\eqref{eq: P_tot(t)_TDH_midstart} for the site-occupation probability $\overline{P}_m(t)$ plotted against $m$ at four different times, with $\lambda=0.5$. Here the particle reaches a stationary state characterized by localization around the two reset sites. Panel (b) shows $\overline{P}_m(t)$ versus $t$ for two different $m$'s, one being the reset site $m=n_r$ and another being a non-reset site $m=-2$, with $\lambda=0.5$. The analytical results given by Eq.~\eqref{eq: P_tot(t)_TDH_midstart} and plotted with lines are compared with numerical results depicted by symbols. The latter results are obtained via numerical implementation of the dynamics on a lattice of $41$ sites and involve averaging over $5\times10^3$ dynamical realizations. The horizontal red dashed lines denote the numerically evaluated stationary-state probability $\overline{P}_m^\mathrm{st}$ given by Eq.~\eqref{eq: P_tot(t)_TDH_midstart-steady-state}.}\label{fig: 7}
\end{figure}

The features seen in the equidistant-reset-site case for the time-independent Hamiltonian~\eqref{eq:H} are also observed for the case when the system is subject to an external forcing field, i.e., for the time-dependent Hamiltonian case given by Eq.~\eqref{eq:time_dep_H}, see Fig.~\ref{fig: 7}. This is an illustration of the robustness of the phenomenon of localization of the TBM particle around the reset sites, which we have demonstrated to hold in a setting involving either a time-independent or a time-dependent Hamiltonian. On another note, the fact that one has a perfect match between analytical and numerical results for both the time-independent and the time-dependent Hamiltonian is a validation of how our general analytical approach outlined in Section~\eqref{sec:general-set-up} is able to tackle efficiently both the situations and enables one to derive explicit analytical results. 

\section{TBM subject to conditional resetting at power-law-distributed times}\label{sec:power-law}

Until now, we have investigated the consequences of conditional resetting in the TBM both in the absence and presence of a forcing field, with the resetting taking place at time intervals that are exponentially distributed. We have observed in all cases a spatial localization of the TBM particle at the two reset sites as a signature of the stationary state that the dynamics attains at long times. We have already addressed the robustness of such a state, by studying the TBM both in the absence and presence of a forcing field. An imminent issue affecting robustness would be to ask if a similar stationary state emerges even when the resetting is done at time intervals that are differently distributed than an exponential. In particular, we would like to consider the case of a power-law distribution. Namely, the time intervals $\tau$ between consecutive resets are distributed according to the distribution
\begin{equation}
    p(\tau) = \frac{\alpha}{{\tau_0(\tau/\tau_0)}^{1+\alpha}}; \quad\text{\hspace{0.03 cm}}\quad \alpha>0; \quad\text{\hspace{0.03 cm}} \tau\in[\tau_0,\infty).\label{eq: power_law_dist}
\end{equation}
Here, $\tau_0$ is the cut-off time scale. Note that for $\alpha < 1$, all moments of $p(\tau)$ are infinite. For $\alpha>1$, the first moment is finite and equals $\langle \tau\rangle =\tau_0\alpha/(\alpha-1)$, while for $\alpha>2$, the second moment also becomes finite: $\langle \tau^2\rangle= \tau_0^2 \alpha/(\alpha - 2)$. By contrast, the previously studied exponential $p(\tau)$, Eq.~\eqref{eq: exp_dist}, has finite mean and variance for all $\lambda>0$. As we will unveil, these facts will have important consequences on the emergence and properties of the stationary state in the case of power-law vis-\`{a}-vis exponential resetting.

\begin{figure}[H]
\includegraphics[width=1\linewidth]{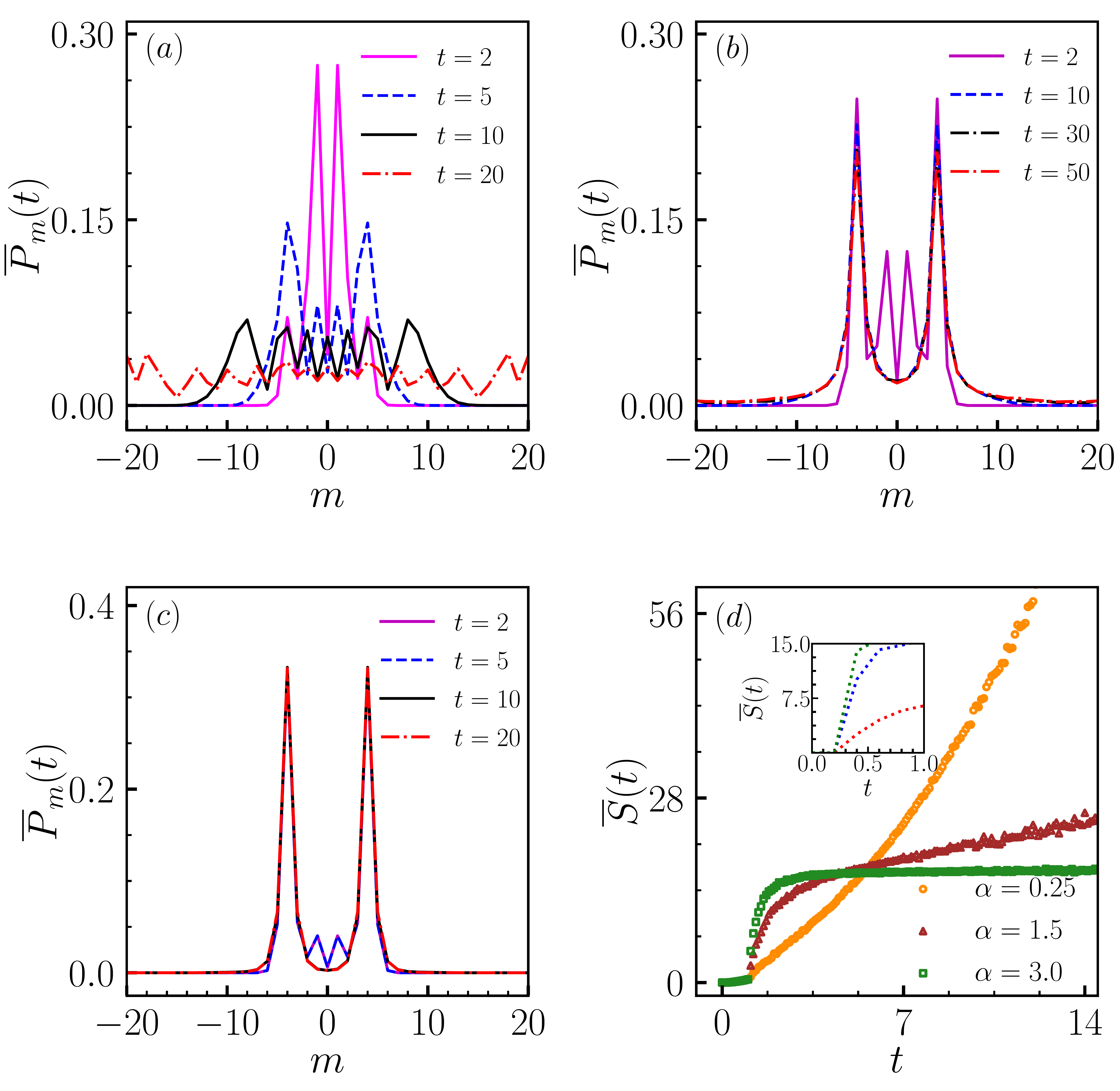}
\caption{For the same TBM lattice as in Fig.~\ref{fig: 4} with conditional resetting being done instead at power-law-distributed time intervals, the figure corresponds to the case of the initial location of the TBM particle being at the origin, $n_0=0$, and with resetting to locations equidistant from the initial location, namely, $n_r=4$ and $n_l=-4$. The numerical results of the site-occupation probability $\overline{P}_m(t)$ versus $m$ at four different times are plotted in panels (a), (b) and (c). In panel (a), for $\alpha=0.25$ ($\alpha<1$), the system does not reach a stationary state and consequently, localization does not take place at the desired sites $n_r$ and $n_l$. By contrast, in panels (b) and (c), for $\alpha=1.5$ ($1<\alpha<2$) and $\alpha=3.0$ ($\alpha>2$), respectively, the system reaches a stationary state at long times (compare data between $t=30$ and $t=50$ in panel (b) and between $t=10$ and $t=20$ in panel (c)). Panel (d) shows numerical results for the MSD $\overline{S}(t)$ versus $t$ for three different values of $\alpha$. The MSD exhibits fast and slow divergence for respectively $\alpha<1$ and $1<\alpha<2$, whereas it reaches a constant value for $\alpha>2$. Although the MSD behaves differently for different values of $\alpha$, they behave similarly up to the cut-off time-scale $\tau_0=1.0$ chosen here,  defined in Eq.~\eqref{eq: power_law_dist}, since the initial state undergoes a reset-free evolution up to time $\tau_0$. As a further validation of the latter fact, the inset shows the numerical results for the MSD versus time with $\tau_0 =0.1$. The presented numerical results are obtained via numerical implementation of the dynamics on a lattice of $41$ sites and involve averaging over $5\times10^3$ dynamical realizations.}\label{fig: 8}
\end{figure}

In the light of the results discussed in Section~\ref{sec:numerics} and displayed in Figs.~\ref{fig: 4},~\ref{fig: 5} and~\ref{fig: 6}, one may argue that the features such as the emergence of a stationary state and the associated localization around the reset sites are what one would have expected anyways, since after all the resetting moves simply do not allow the particle to travel very far from the reset sites. Consequently, it should be no wonder that one would have localization in the presence of resetting, whose very absence makes the bare TBM particle travel to further and further distances with the passage of time, thereby preventing the existence of a stationary state with a time-independent MSD at long times. Resetting at power-law times provides a crucial counter to such a line of reasoning. Indeed, the results depicted in Fig.~\ref{fig: 8}(a) show that when the exponent $\alpha$ is smaller than unity, there is no stationary state emerging and one has an ever-expanding occupation probability in time despite the resetting events trying to confine the particle. Concomitant with such a behavior is that of the MSD depicted in panel (d), which may be seen to grow unbounded in time for $0<\alpha<1$. On the contrary, the data for $\alpha>2$ do put in evidence that resetting is able to confine the particle, which therefore has a localized stationary state (compare the plots in panels (c) and (d)). The case of $\alpha$ lying between $1$ and $2$ is particularly interesting (and telling for the underlying interplay of time scales, see below), in that here, the site-occupation probability does have a stationary state and yet the MSD continues to grow unbounded in time (compare the plots in panels (b) and (d)). We thus conclude that the attainment of a stationary state in presence of conditional resetting
is not always evident and obvious, and depends crucially on
the probability distribution of the intervals of time at which resetting is implemented. 

The reason one does not have a stationary state in the case of power-law resetting with $0<\alpha<1$ may be traced back to some crucial features of the power-law distribution~\eqref{eq: power_law_dist}, namely, those related to the distribution having finite mean and variance. Indeed, for $0<\alpha<1$, when the mean is infinite, it so happens that there is a significant number of dynamical realizations in which the TBM particle gets very far from the initial location in time before it gets reset, and this is because with significant probability one may sample reset time intervals that are very large, larger than the time scale $1/\gamma$ for the TBM particle to traverse a finite distance over the lattice. Such a scenario is obviously not tenable when the mean is finite, which is really the case for $\alpha>1$ and which accounts for the existence of a stationary state for $\alpha>1$. What then is the reason for a diverging MSD for $1<\alpha<2$ when the site-occupation probability becomes stationary at long times? This is attributed to a diverging variance of the power-law distribution~\eqref{eq: power_law_dist} in this range of values of $\alpha$. Similar effects have been observed and analyzed, albeit in the classical resetting scenario, in Ref.~\cite{PhysRevE.93.060102}.

\section{Conclusions}
\label{sec:conclusion}
In this work, we developed a time-dependent and conditional resetting protocol as a model for investigating the unitary evolution of quantum systems subjected to external interventions at random times. Applying this protocol to a tight-binding model (TBM) involving a quantum particle undergoing hopping to nearest-neighbour sites on a one-dimensional open lattice, we provided both analytical and numerical evidence for existence of stationary states characterized by the localization of the TBM particle around the reset sites. This is in stark contrast with the bare TBM dynamics, which does not induce any such stationary states characterized by localization.

In the course of presenting things, we first studied the effect of our conditional resetting protocol on the time-independent TBM Hamiltonian dynamics, with the resetting time intervals chosen from an exponential distribution. The dynamics comprises two recurring key events: unitary evolution for a random time interval followed by a conditional reset of the TBM particle to two reset locations, conditioned on the location of the particle at the time instant of reset. The choice of initial location and the location of the reset sites played crucial roles in dictating the nature of the stationary-state site-occupation probability of the TBM particle. Specifically, reset sites equidistant from the initial location taken to
be at the origin resulted in symmetric probability profile about the origin. Even when the initial location of the particle was chosen to be any site other than the origin, with symmetrically disposed reset sites around the origin, our protocol led at long times to an eventual symmetry in the site-occupation probability about the origin. This is in conformity with the intuition of a stationary state being independent of the choice of the initial condition. Nevertheless, in the second case and at small times, an asymmetry in the occupation probability emerged because of the memory, carried over from past events, due to the time-dependent and conditional structure of the reset protocol. Considering the reset locations asymmetrically disposed around the origin led to the result that, even at large times, the site-occupation probability has an asymmetry about the origin and becomes highly peaked around one reset location than the other. Interestingly, in this case, while there was no inherent bias in the reset-free dynamics, the protocol itself induces an effective bias leading to enhanced localization around one of the reset sites. We extended our approach to include the effects of a periodic forcing within the TBM. Analytical derivation of the resulting stationary state site-occupation probability showcased the versatility of our analytical approach with respect to either  time-independent or time-dependent Hamiltonians.

Further to the aforementioned analysis, we examined the impact of the choice of the resetting interval distribution on the emergence of the stationary state. Notably, stationary states do not appear when the resetting intervals followed a power-law distribution for small values of the corresponding exponent, which implies that it is not always true that resetting induces a stationary state. As a result, contrary to the case of exponential resetting for which the mean-squared displacement (MSD) of the TBM particle about the initial location always reaches a stationary value in time, the MSD may diverge in time when the resetting time intervals are chosen from a power-law distribution. 

As stated in the Introduction, our protocol of conditional resetting may be exploited to explore consequences emerging in non-Markovian dynamics of open quantum systems. Here, we propose a possible experimental platform that may be utilized efficiently to implement and observe consequences of conditional resetting. To this end, we consider the prototypical atomic-physics set-up of level populations under the influence of an incident light field. As one may expect, the light field would induce oscillations among the populations in the various levels. One may engineer a system in which there exist two metastable states close to two selected levels in the system. In order to implement conditional resetting at random times, population oscillations would be allowed to continue for a random time after which a pair of light fields of stronger intensity than the first light field would be turned on, which has the frequency spectrum to excite populations around the two selected levels to the respective metastable states. Such an excitation would be conditional on the existing population size of the levels prior to turning on the pair of second light fields. The light field pair would then be turned off, following which there would be de-excitation to the two selected levels from the respective metastable states, thereby implying conditional resetting to the two levels. Subsequently, effect of the first light field (which is kept on all throughout the experiment) would effect redistribution of the population among the different levels and consequent population oscillations. The latter would continue for another random time, when again the light field pair would be turned on, and so on. In recent times, similar set-ups involving a radio frequency pulse and a Raman pulse have been employed to implement experimentally projective measurements at random times using a Bose–Einstein condensate of $^{87}\mathrm{Rb}$ produced in a magnetic micro-trap realised with an atom chip~\cite{gherardini2017ergodicity}.

In terms of future perspectives, it would be interesting to apply our time-dependent and conditional resetting formalism by incorporating a bias in the bare TBM dynamics, details of which will be reported elsewhere. Introducing periodicity in the TBM lattice would also be interesting in studying the nature of the ensuing stationary-state properties.

\section*{Acknowledgements}We thank Soumya Kanti Pal, Vaibhav Prabhudesai and C L Sriram for useful discussions. We gratefully acknowledge generous
allocation of computing resources by the Department of
Theoretical Physics (DTP) of the Tata Institute of Fundamental Research (TIFR), and related technical assistance from Kapil Ghadiali and Ajay Salve. AA acknowledges useful discussions with Shatanik Bhattacharya, Rupak Majumder and Mrinal Kanti Sarkar.~SG acknowledges support from the Science and Engineering Research Board (SERB), India under SERB-CRG scheme Grant No. CRG/2020/000596. He also thanks ICTP–Abdus Salam International Centre for Theoretical Physics, Trieste, Italy, for support under its Regular Associateship scheme. SG is grateful to the SISSA visiting scientist program and ICTP, Trieste for financial support towards his visit in July 2023 when this paper was being finalized. SG would also like to thank the Isaac Newton Institute for Mathematical Sciences, Cambridge, for support and hospitality during the programme ``Mathematics of movement: An interdisciplinary approach to mutual challenges in animal ecology and cell biology," where work on this paper was completed. This work was supported by EPSRC grant no EP/R014604/1.

\appendix
\section{Derivation of Eq.~\eqref{eq:Lindblad-ours}}
\label{sec:app0}
From Eq.~\eqref{eq:average-rhot_tbh} and considering the case of exponential $p(\tau)$, Eq.~\eqref{eq: exp_dist}, the average density operator at time $t$ may be written as 
\begin{align}
\overline{\rho}(t)= \sum_{\alpha=0}^{\infty} \overline{\rho}^{(\alpha)}(t), 
\label{eq:rho-evolution-rho-sum}
\end{align}
with   
\begin{align}
&\overline{\rho}^{(0)}(t) \equiv F(t) \mathrm{e}_+^{-\mathrm{i} \int_0^{t} \mathrm{d}t' \mathcal{L}(t')}\rho(0),\\
&\overline{\rho}^{(\alpha)}(t) \equiv \int_{0}^{t} \mathrm{d}{t_\alpha}~\mathrm{e}_+^{-\mathrm{i}\int_{t_\alpha}^{t} \mathrm{d}t'\mathcal{L}(t')} F(t-t_\alpha)G_\alpha(t_\alpha);~\alpha \ge 1, 
\end{align}
where we have $p(t)=\lambda \mathrm{e}^{-\lambda t}=\lambda F(t)$, and 
\begin{align}
&G_1(t) \equiv T(t)~ p(t)~ \mathrm{e}_+^{-\mathrm{i}\int_0^{t} \mathrm{d}t'~\mathcal{L}(t')} \rho(0), \\
&G_\alpha(t) \equiv \int_{0}^{t} \mathrm{d}{t_{\alpha-1}}  T(t-t_{\alpha-1}) p(t-t_{\alpha-1})\nonumber \\
&~~~~~~~~\times \mathrm{e}_+^{-\mathrm{i}\int_{t_{\alpha-1}}^{t} \mathrm{d}t'~\mathcal{L}(t')} G_{\alpha-1}(t_{\alpha-1});~\alpha\ge 2.
\end{align}
Differentiating Eq.~\eqref{eq:rho-evolution-rho-sum} with respect to $t$, one obtains on using $F(0)=1$ that
\begin{align}
&\frac{\mathrm{d}\overline{\rho}^{(0)}(t)}{\mathrm{d} t} =- \mathrm{i} \mathcal{L}(t) \overline{\rho}^{(0)}(t)- \lambda \overline{\rho}^{(0)}(t), \\
&\frac{\mathrm{d}\overline{\rho}^{(\alpha)}(t)}{\mathrm{d} t}= - \mathrm{i} \mathcal{L}(t) ~\overline{\rho}^{(\alpha)}(t) + F(0) G_\alpha(t)\nonumber \\
&~~~~+ \int_{0}^{t} \mathrm{d}t_\alpha~\frac{\mathrm{d}}{\mathrm{d} t}F(t-t_\alpha)  ~\mathrm{e}_+^{-\mathrm{i} \int_{t_\alpha}^{t} \mathrm{d}t'\mathcal{L}(t')} G_\alpha(t_\alpha)\\
&~~~~=- \mathrm{i} \mathcal{L}(t)\overline{\rho}^{(\alpha)}(t) + G_\alpha(t)-\lambda \overline{\rho}^{(\alpha)}(t);~\alpha \ge 1.
\end{align}
Summing over all $\alpha$ terms yields
\begin{align}
&\frac{\mathrm{d}\overline{\rho}(t)}{\mathrm{d} t}=-\mathrm{i} \mathcal{L}(t)\overline{\rho}(t)+ \lambda \sum_{\alpha=1}^{\infty} H_{\alpha}(t)-\lambda \overline{\rho}(t),
\label{eq:lindblad-app0}
\end{align}
with $H_\alpha(t)=G_\alpha(t)/\lambda;~\alpha \ge 1$. We thus obtain Eq.~\eqref{eq:Lindblad-ours} of the main text.

For the case of a time-independent $T$, we get
\begin{align}
&G_1(t)=\lambda TF(t)\mathrm{e}_+^{-\mathrm{i}\int_0^{t} \mathrm{d}t'~\mathcal{L}(t')} \rho(0)=\lambda T\overline{\rho}^{(0)}(t), 
\end{align}
while for $\alpha \ge 2$, we have
\begin{align}
&G_\alpha(t)\nonumber \\
&= \lambda T \int_{0}^{t} \mathrm{d}{t_{\alpha-1}}  F(t-t_{\alpha-1})\mathrm{e}_+^{-\mathrm{i}\int_{t_{\alpha-1}}^{t} \mathrm{d}t'~\mathcal{L}(t')} G_{\alpha-1}(t_{\alpha-1})\nonumber \\
&=\lambda T\overline{\rho}^{(\alpha-1)}(t).
\end{align}
Consequently, we have
\begin{align}
\lambda \sum_{\alpha=1}^\infty H_\alpha(t)=\sum_{\alpha=1}^\infty G_\alpha(t)=\lambda T\overline{\rho}(t),
\end{align}
and Eq.~\eqref{eq:lindblad-app0} reduces to 
\begin{align}
&\frac{\mathrm{d}\overline{\rho}(t)}{\mathrm{d} t}=-\mathrm{i} \mathcal{L}(t)\overline{\rho}(t)+ \lambda T\overline{\rho}(t) -\lambda \overline{\rho}(t),
\label{eq:lindblad-app1}
\end{align}
which is the result derived in Ref.~\cite{Debraj:2022-2} (see also the discussion around Eq.~\eqref{eq:Lindblad-1} in the Introduction).

\section{A few useful identities related to superoperators}
\label{sec:app1}
Here, we will give a brief overview of the notion of a  superoperator, which we have used extensively in the main text. The concept of a superoperator bears resemblance to that of an operator acting on state vectors as commonly encountered in quantum mechanics. Just as an operator acting on a state vector yields another state vector, a superoperator acting on an operator yields a new operator. We now discuss the notion of a superoperator in the context of the TBM, for which a complete set of basis states is formed by the Wannier states $\{|n\rangle\}$. Let $\mathcal{O}$ be a superoperator, whose operation on an operator $A$ is defined via respective matrix elements, as~\cite{Debraj:2022}
\begin{align}
   \langle n_1| \mathcal{O} A |n_2 \rangle=\sum_{n_3 n_4} ( n_1 n_2|\mathcal{O} | n_3 n_4) \langle n_3| A| n_4 \rangle, \label{eq: super operation matrix element}
\end{align}
with $|mn)\equiv |m\rangle \otimes |n\rangle$, and wherein the `matrix elements' of the superoperator have been labeled by four indices, just as those of an ordinary operator are labeled by two indices. The superoperator lives in a product Hilbert space, the dimension of which is the square of the space of the associated operator. One has the closure property for the two-indexed states as
$\sum_{m,n}|mn)(mn| = I$. The familiar Liouville operator $\mathcal{L}$ is a superoperator. It has the well-known property that acting on any operator $A$, it produces the commutation between $H$ and $A$, with $H$ being the Hamiltonian of the system:
\begin{align}
    \mathcal{L} A = [H,A].
\end{align}
Following Eq.~\eqref{eq: super operation matrix element}, it is easily checked that
\begin{align}
  ( n_1 n_2|\mathcal{L} | n_3 n_4)=  \langle n_1 | H | n_3 \rangle \delta_{n_2 n_4} - \langle n_4 | H | n_2 \rangle \delta_{n_1 n_3}.\label{matrix element of L}   
\end{align}
Now, the time-evolution of a density operator $\rho(t)$ under a time-independent Hamiltonian is given by
\begin{align}
\rho(t)=\mathrm{e}^{-\mathrm{i} Ht}\rho(0)\mathrm{e}^{\mathrm{i} Ht}=\mathrm{e}^{-\mathrm{i}{\mathcal L}t}\rho(0).
\label{eq:liouville-eqn}
\end{align}
We may now calculate the matrix elements of $\rho(t)$, as
\begin{align}
\langle n_1|\rho(t)|n_2\rangle&=\sum_{n_3 n_4}(n_1 n_2|\mathrm{e}^{-{\mathrm{i}}{\mathcal L}t}|n_3 n_4)\langle n_3|\rho(0)|n_4\rangle\nonumber\\
&=\sum_{n_3 n_4}\langle n_1|\mathrm{e}^{-{\mathrm{i}}Ht} |n_3\rangle  \langle n_4|\mathrm{e}^{{\mathrm{i}}Ht}|n_2\rangle \langle n_3|\rho(0)|n_4\rangle,
\label{eq:rho-matrix-element}
\end{align}
yielding
\begin{align}
(n_1 n_2|\mathrm{e}^{-{\mathrm{i}}{\mathcal L}t}|n_3 n_4) = \langle n_1|\mathrm{e}^{-{\mathrm{i}}Ht} |n_3\rangle  \langle n_4|\mathrm{e}^{{\mathrm{i}}Ht}|n_2\rangle. \label{eq:iLt-elements}
\end{align}

\section{Derivation of Eq.~(41) of the main text}
\label{sec:app2}
In this appendix, we derive Eq.~\eqref{eq: Pm1(t)} of the main text. We start with the second term of the series in Eq.~{\eqref{eq: probability series terms}}, which may be evaluated by using the matrix element of the superoperator $T$ given by expression~\eqref{eq: T1(t)}:
\begin{align}
&\widetilde{\overline{P}}_m^{(1)}(s)\nonumber\\
    &=\lambda \langle m |\widetilde{U}_0(s)\widetilde{T}'(s')\rho(0)|m \rangle \nonumber\\
    &=\lambda\int_0^\infty \mathrm{d} \tau \:\mathrm{e}^{-(s+\lambda) \tau} \sum_{\substack{n_1,n_2,n_3,\\n_4,n_5,n_6}}(m m| \widetilde{U}_0(s)| n_1 n_2)\nonumber\\ 
    &\times (n_1 n_2| T(\tau)|n_3 n_4) (n_3 n_4|\mathrm{e}^{-\mathrm{i}\mathcal{L} \tau} | n_5 n_6) \langle n_5 | \rho(0)| n_6 \rangle \nonumber\\
    &=\lambda\int_0^\infty \mathrm{d}\tau\: \mathrm{e}^{-(s+\lambda) \tau}\! \sum_{\substack{n_1,n_2,n_3}}(m m| \widetilde{U}_0(s)| n_1 n_2)\Big(\delta_{n_1 n_r}\delta_{n_2 n_r} \nonumber\\
    & \Omega_{n_r}^{n_0}(\gamma\tau)+\delta_{n_1 n_l}\delta_{n_2 n_l} \Omega_{n_l}^{n_0}(\gamma\tau)\Big)(n_3 n_3| \mathrm{e}^{-\mathrm{i}\mathcal{L} \tau}| n_0 n_0) \nonumber
    \\
    &=\lambda\int_0^\infty \mathrm{d}\tau\; \mathrm{e}^{-(s+\lambda) \tau}\dfrac{1}{( 2 \pi)^2}\int_{-\pi}^\pi \mathrm{d}k \int_{-\pi}^\pi \mathrm{d}k\dfrac{1}{(s+\lambda)I+ \mathrm{i} \Gamma_{k k'}}
    \nonumber\\
    & \times(\mathrm{e}^{\mathrm{i}(m-n_r)(k-k')}\Omega_{n_r}^{n_0}(\gamma \tau)+\mathrm{e}^{\mathrm{i}(m-n_l)(k-k')}\Omega_{n_l}^{n_0}(\gamma\tau) ).\label{eq: Pm1}
\end{align}
 Here, in the second step, we have used Eq.~\eqref{eq: super operation matrix element}, while the fourth step uses the result $\sum_{n_3}(n_3 n_3|\mathrm{e}^{-\mathrm{i}\mathcal{L}\tau}| n_0 n_0)=1$, which may be shown as follows: 
 \begin{align}  
 &\sum_{n_3}(n_3 n_3|\mathrm{e}^{-\mathrm{i}\mathcal{L}\tau}| n_0 n_0)\nonumber\\
 &= \sum_{n_3}\langle n_3 | \mathrm{e}^{- \mathrm{i} H \tau} |n_0\rangle \langle n_0| \mathrm{e}^{ \mathrm{i} H \tau} |n_3\rangle \nonumber\\
 &=\sum_{n_3}\dfrac{1}{( 2 \pi)^2}\int_{-\pi}^\pi \mathrm{d}k \int_{-\pi}^\pi \mathrm{d}k'\:\mathrm{e}^{\mathrm{i} (n_3-n_0)(k-k')} \mathrm{e}^{-\mathrm{i} \Gamma_{kk'} \tau}\nonumber\\
 &=\sum_{n_3} \; J^2_{|n_3-n_0|}(\gamma \tau) = 1,
\end{align}
where, in obtaining the second step, we have used Eqs.~\eqref{eq:bare-TBM-result} and~\eqref{eq:Bloch-states}, while in arriving at the third step, we have used Eq.~\eqref{free_prop}.
Using the Laplace convolution theorem, we obtain from Eq.~\eqref{eq: Pm1} on use of Eq.~\eqref{free_prop} the corresponding result in the time domain: 
\begin{align}
\overline{P}_m^{(1)}(t)
   &= \lambda \mathrm{e}^{-\lambda t}\int_0^t \mathrm{d}t_1  ( J^2_{|m-n_r|}(\gamma (t-t_1) ) \Omega_{n_r}^{n_0}(\gamma t_1)\nonumber\\
   &+  J^2_{|m-n_l|}(\gamma (t-t_1))
   \Omega_{n_l}^{n_0}(\gamma t_1)),
\end{align}
which is Eq.~\eqref{eq: Pm1(t)} of the main text.

\section{Derivation of Eq.~(42) of the main text}
\label{sec:app3}
This appendix is devoted to a derivation of Eq.~\eqref{eq: Pm2(t)} of the main text.
The third term of the series in Eq.~\eqref{eq: probability series terms} can be calculated by using the matrix element of superoperator $T$ given by the expressions \eqref{eq: T1(t)} and \eqref{eq: T2(t)}, as 
\begin{widetext}
    \begin{align}       &\widetilde{\overline{P}}_m^{(2)}(s)\nonumber\\
        &=\lambda^2 \langle m |\widetilde{U}_0(s)\Tilde{T}(s')\Tilde{T}(s') \rho(0)|m \rangle\nonumber\\
        &=\lambda^2 \int_0^\infty \mathrm{d}\tau_2 \int_0^\infty \mathrm{d}\tau_1 \;\mathrm{e}^{-(s+\lambda)(\tau_1+\tau_2)} \sum_{n_i's}(m m| \widetilde{U}_0(s)| n_1 n_2)(n_1 n_2| T(\tau_2)|n_3 n_4)(n_3 n_4| \mathrm{e}^{-\mathrm{i}\mathcal{L} \tau_2}| n_5 n_6)
       \nonumber\\
       & \times(n_5 n_6| T(\tau_1)|n_7  n_8)(n_7 n_8 |\mathrm{e}^{-\mathrm{i}\mathcal{L} \tau_1} | n_9 n_{10})\langle n_9 | \rho(0)| n_{10}\rangle \nonumber\\
       &=\lambda^2 \int_0^\infty \mathrm{d}\tau_2\int_0^\infty \mathrm{d}\tau_1\;\mathrm{e}^{-(s+\lambda)(\tau_1+\tau_2)}\sum_{n_1 n_2} (m m | \widetilde{U}_0(s) | n_1 n_2) (\delta_{n_1 n_r} \delta_{n_2 n_r} (\Omega_{n_r}^{n_l}(\gamma\tau_2) \Omega_{n_l}^{n_0}(\gamma\tau_1) + \Omega_{n_r}^{n_r}(\gamma\tau_2) \Omega_{n_r}^{n_0}(\gamma\tau_1))\nonumber\\    
        &+\delta_{n_1 n_l}\delta_{n_2 n_l}(
          \Omega_{n_l}^{n_l}(\gamma\tau_2)\Omega_{n_l}^{n_0}(\gamma\tau_1)+  \Omega_{n_l}^{n_r}(\gamma\tau_2)\Omega_{n_r}^{n_0}(\gamma\tau_1)))
       \Bigl[\sum_{n_3}(n_3 n_3 |\mathrm{e}^{-\mathrm{i}\mathcal{L} \tau_2}| n_r n_r) \Omega_{n_r}^{n_0}(\gamma\tau_1)+\sum_{n_3}(n_3 n_3 |\mathrm{e}^{-\mathrm{i}\mathcal{L} \tau_2}|n_l n_l) \Omega_{n_l}^{n_0}(\gamma\tau_1)\Bigr]\nonumber\\
        &\times\sum_{n_7}(n_7 n_7|\mathrm{e}^{-\mathrm{i}\mathcal{L} \tau_1}| n_0 n_0 )\nonumber\\
      &= \dfrac{\lambda^2}{( 2 \pi)^2}\int_0^\infty \mathrm{d}\tau_2\int_0^\infty\mathrm{d}\tau_1\; \mathrm{e}^{-(s+\lambda)(\tau_1+\tau_2)}\int_{-\pi}^\pi \mathrm{d}k \int_{-\pi}^\pi  \mathrm{d}k' \dfrac{1}{(s+\lambda)I+\mathrm{i}\Gamma_{k k'}}
      \Bigl[ \mathrm{e}^{ \mathrm{i}(m-n_r)(k-k')}  (\Omega_{n_r}^{n_l}(\gamma\tau_2) \Omega_{n_l}^{n_0}(\gamma\tau_1)
      \nonumber\\
      &+\Omega_{n_r}^{n_r}(\gamma\tau_2)\Omega_{n_r}^{n_0}(\gamma\tau_1))+ \mathrm{e}^{ \mathrm{i}(m-n_l)(k-k')} (\Omega_{n_l}^{n_l}(\gamma\tau_2) \Omega_{n_l}^{n_0}(\gamma\tau_1)+\Omega_{n_l}^{n_r}(\gamma\tau_2)\Omega_{n_r}^{n_0}(\gamma\tau_1))\Bigr],\label{eq: Pm_2(s)}
    \end{align}
\end{widetext}
where the fourth step is obtained using the results $\sum_{n_3}(n_3n_3|\mathrm{e}^{-\mathrm{i}\mathcal{L}\tau_2}|n_0n_0)=1$ , $\sum_{n_7}(n_7n_7|\mathrm{e}^{-\mathrm{i}\mathcal{L}\tau_1}|n_0n_0)=1$, and Eq.~\eqref{eq:norm-Q}. Consequently, in the time domain, on using Eq.~\eqref{free_prop} in Eq.~\eqref{eq: Pm_2(s)}, we obtain $\overline{P}_m^{(2)}(t)$ as
\begin{widetext} 
\begin{align}
\overline{P}_m^{(2)}(t)
    &=\lambda^2 \mathrm{e}^{-\lambda t } \int_0^t \mathrm{d} t_2 \int_0^{t_2} \mathrm{d} t_1 \Big[J_{|m-n_r|}^2(\gamma (t-t_2))\left(\Omega_{n_r}^{n_r}(\gamma(t_2-t_1)) 
     \Omega_{n_r}^{n_0}(\gamma t_1)+\Omega_{n_r}^{n_l}(\gamma(t_2-t_1))\Omega_{n_l}^{n_0}(\gamma t_1)\right)\nonumber\\
    &+J_{|m-n_l|}^2(\gamma (t-t_2) ) \left(\Omega_{n_l}^{n_r}(\gamma(t_2-t_1))\Omega_{n_r}^{n_0}(\gamma t_1)+\Omega_{n_l}^{n_l}(\gamma(t_2-t_1))\Omega_{n_l}^{n_0}(\gamma t_1)\right)\Big],
   \end{align}
   \end{widetext}
which is Eq.~\eqref{eq: Pm2(t)} of the main text.


\bibliography{ref_paper}

\end{document}